\documentclass[twocolumn, twocolappendix, numberedappendix, appendixfloats]{openjournal}

\usepackage{xcolor}
\usepackage{textgreek}
\usepackage[utf8]{inputenc}
\usepackage[english]{babel}

\usepackage{hyperref}
\hypersetup{
    unicode,
    colorlinks=true,
    linkcolor=linkcolor,
    citecolor=linkcolor,
    filecolor=linkcolor,
    urlcolor=linkcolor,
}
\usepackage{color,colortbl}
\definecolor{linkcolor}{rgb}{0.0,0.3,0.5}
\usepackage{tensind}
\tensordelimiter{?}
\DeclareGraphicsExtensions{.bmp,.png,.jpg,.pdf}
\usepackage{verbatim}
\usepackage[normalem]{ulem}
\usepackage{orcidlink}
\usepackage{soul}
\usepackage{enumitem}
\usepackage{booktabs}
\usepackage{ragged2e}

\graphicspath{ {./figures/} }

\usepackage{ifthen}
\newcommand{\figplaceholder}[1]{%
  \setlength{\fboxsep}{0pt}%
  \framebox[\linewidth]{\parbox[c][0.55\linewidth][c]{\linewidth}{\centering\vspace{1ex}\textit{Placeholder for figure:}\\\texttt{#1}}}%
}
\newcommand{\safeincludegraphics}[2][]{%
  \IfFileExists{#2}{\includegraphics[#1]{#2}}{\figplaceholder{#2}}%
}

\usepackage[T1]{fontenc}

\DeclareRobustCommand{\VAN}[3]{#2}
\let\VANthebibliography\thebibliography
\def\thebibliography{\DeclareRobustCommand{\VAN}[3]{##3}\VANthebibliography}

\usepackage{graphicx}	
\usepackage{adjustbox}  
\usepackage{amsmath}	
\usepackage{amssymb}	
\usepackage{wasysym}    
\usepackage{bm}         
\usepackage{xspace}
\usepackage{soul}
\usepackage{multirow}
\makeatletter 
  \patchcmd{\NAT@citex}
    {\@citea\NAT@hyper@{%
      \NAT@nmfmt{\NAT@nm}%
      \hyper@natlinkbreak{\NAT@aysep\NAT@spacechar}{\@citeb\@extra@b@citeb}%
      \NAT@date}}
    {\@citea\NAT@nmfmt{\NAT@nm}%
    \NAT@aysep\NAT@spacechar\NAT@hyper@{\NAT@date}}{}{}

  \patchcmd{\NAT@citex}
    {\@citea\NAT@hyper@{%
      \NAT@nmfmt{\NAT@nm}%
      \hyper@natlinkbreak{\NAT@spacechar\NAT@@open\if*#1*\else#1\NAT@spacechar\fi}%
        {\@citeb\@extra@b@citeb}%
      \NAT@date}}
    {\@citea\NAT@nmfmt{\NAT@nm}%
    \NAT@spacechar\NAT@@open\if*#1*\else#1\NAT@spacechar\fi\NAT@hyper@{\NAT@date}}
    {}{}
\makeatother

\newcommand\Msun{\text{M}_{\astrosun}}

\DeclareRobustCommand{\HII}{{\text{H}\,\textsc{ii}}\xspace} 
\DeclareRobustCommand{\HeII}{{\text{He}\,\textsc{ii}}\xspace} 
\DeclareRobustCommand{\HeIII}{{\text{He}\,\textsc{iii}}\xspace} 
\DeclareRobustCommand{\thesan}{\mbox{\textsc{thesan}}\xspace}
\DeclareRobustCommand{\thesanzoom}{\mbox{\textsc{thesan-zoom}}\xspace}
\DeclareRobustCommand{\lumina}{\mbox{\textsc{lumina}}\xspace}
\DeclareRobustCommand{\thesanone}{\mbox{\textsc{thesan-1}}\xspace}

\DeclareRobustCommand{\thesanhr}{\mbox{\textsc{thesan-hr}}\xspace}
\DeclareRobustCommand{\arepo}{\mbox{\textsc{arepo}}\xspace}
\DeclareRobustCommand{\areport}{\mbox{\textsc{arepo-rt}}\xspace}

\newcommand\orcid[1]{\href{http://orcid.org/#1}{\adjustbox{trim={-.15\width 0 -.15\width 0\height},clip}{\includegraphics[height=9pt]{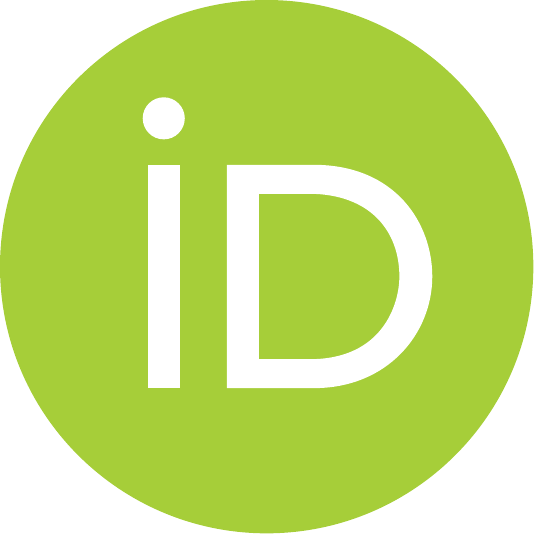}}}}
\newcommand{\appropto}{\mathrel{\vcenter{
  \offinterlineskip\halign{\hfil$##$\cr
    \propto\cr\noalign{\kern2pt}\sim\cr\noalign{\kern-2pt}}}}}

\begin{document}
\title{The Lumina Project: CMB Optical Depth Fluctuations from Patchy Reionization}

\author{Aaron~Smith\orcid{0000-0002-2838-9033}$^{1\,\star}$}
\email[$^\star$ E-mail: \href{mailto:asmith@utdallas.edu}{asmith@utdallas.edu}]{}
\author{Oliver~Zier\orcid{0000-0003-1811-8915}$^{2}$}
\author{Rahul~Kannan\orcid{0000-0001-6092-2187}$^{3}$}
\author{Xuejian~Shen\orcid{0000-0002-6196-823X}$^{4}$}
\author{Rongrong~Liu\orcid{0000-0003-0685-3525}$^{2}$}
\author{Mark~Vogelsberger\orcid{0000-0001-8593-7692}$^{4}$}
\author{Volker~Springel\orcid{0000-0001-5976-4599}$^{5}$}
\author{R\"udiger~Pakmor\orcid{0000-0003-3308-2420}$^{5}$}
\author{Sonja~M.~Koehler\orcid{0009-0008-0814-3328}$^{2}$}
\author{Lars~Hernquist\orcid{0000-0001-6950-1629}$^{2}$}
\author{Meredith~Neyer\orcid{0000-0002-9205-9717}$^{4}$}
\affiliation{$^{1}$Department of Physics, The University of Texas at Dallas, Richardson, Texas 75080, USA}
\affiliation{$^{2}$Center for Astrophysics $|$ Harvard $\&$ Smithsonian, 60 Garden Street, Cambridge, MA 02138, USA}
\affiliation{$^{3}$Department of Physics and Astronomy, York University, 4700 Keele Street, Toronto, ON M3J 1P3, Canada}
\affiliation{$^{4}$Department of Physics, Kavli Institute for Astrophysics and Space Research, Massachusetts Institute of Technology, Cambridge, MA 02139, USA}
\affiliation{$^{5}$Max Planck Institute for Astrophysics, Karl-Schwarzschild-Str.\ 1, D-85741 Garching, Germany}

\begin{abstract}
  Patchy reionization couples the ionized-bubble morphology to the underlying density field, making the CMB Thomson optical depth sensitive to both the global ionization history and anisotropic fluctuations on the sky. Using the large-volume radiation-hydrodynamical \lumina simulation, we compute $\tau_\text{CMB}$ in two ways: (\textit{i}) from global volume- and mass-weighted ionization histories, and (\textit{ii}) from explicit line-of-sight integrations through on-the-fly light cones. We find that the sightline-averaged optical depth in the light cone, $\langle \tau_\text{LOS} \rangle = 0.0550$, exceeds the value inferred from a global volume-weighted history, $\tau_{\text{CMB},V} = 0.0515$, by $\approx 7\%$. This enhancement is largely captured by the global mass-weighted prediction, $\tau_{\text{CMB},m} = 0.0544$, indicating that precision comparisons to CMB optical-depth constraints should use mass-weighted electron fractions or explicit light-cone integration rather than volume-weighted ionized fractions alone. The excess optical depth accumulates primarily near $z_\text{LOS} = 8.0^{+1.9}_{-1.3}$, where the combination of high physical density and strong ionization-field patchiness is greatest. The resulting $\tau_\text{LOS}$ field is non-Gaussian and exhibits $\gtrsim 5\%$ sightline-to-sightline scatter, with fluctuations tracing rare early-ionized overdensities and large-scale structure. Coarse-graining experiments show that smoothing the ionization field on $\gtrsim 3$\,cMpc scales suppresses the density--ionization correlation and biases $\tau_\text{CMB}$ low relative to the resolved calculation. Finally, angular power spectra and real-space correlation functions decomposed into \HII, \HeII, and \HeIII auto- and cross-contributions reveal scale-dependent departures from simple hydrogen--helium co-tracing and evolving characteristic scales with redshift.
\end{abstract}


\maketitle


\section{Introduction}

Cosmic reionization is the last major phase transition of the intergalactic medium (IGM), transforming a predominantly neutral Universe into a highly ionized plasma under the combined influence of the first generations of galaxies and quasars. Its timing, duration, and morphology encode the emergence and clustering of ionizing sources, photon propagation through an inhomogeneous IGM, and feedback-regulated star formation and black-hole growth \citep{McQuinn2016, Wise2019, Gnedin2022}. Establishing a self-consistent picture of reionization is therefore central to connecting high-redshift galaxy formation to large-scale cosmological observables \citep{Dayal2018, Robertson2022}.

A particularly important constraint on the Epoch of Reionization (EoR) is the Thomson scattering optical depth to the cosmic microwave background (CMB), denoted as $\tau_\text{CMB}$, which sets the overall amount of CMB screening and strongly impacts the amplitude of large-angle E-mode polarization generated during reionization \citep{HuDodelson2002}. As $\tau_\text{CMB}$ represents an integral over the post-recombination free-electron density along the line of sight, it provides a robust global constraint on the amount of ionization at high redshift even when the detailed history is uncertain. In an inhomogeneous, inside-out reionization, the electron density $\langle n_e \rangle$ depends on correlations between density and ionization, so $\tau_\text{CMB}$ predictions based on volume-weighted ionized fractions can be biased even when the volume-averaged history is correct. The inferred value has evolved as analyses of large-scale polarization have improved \citep{Spergel2003, Alvarez2006, Lewis2006, Hinshaw2013, Planck2016, Planck2020}, and $\tau_\text{CMB}$ is routinely used to calibrate simulations and to interpret reionization-model parameter spaces \citep{Iliev2007, Trac2007, Greig2017}. However, predicting CMB observables from first-principles simulations is nontrivial. For example, $\tau_\text{CMB}$ fluctuations are sensitive to the correlation between the overall reionization topology and the underlying matter distribution, which requires both large volumes for statistically representative samples and sufficient resolution to capture ionized bubble structure self-consistently with the local and environmental density field.

Observational constraints on the reionization history have improved significantly in the last few decades, but independent EoR probes still leave room for different detailed histories. Quasar absorption spectroscopy constrains the late stages of hydrogen reionization and, when combined with a CMB optical depth measurement, informs both the duration of reionization and the character of the ionizing sources \citep{Sharma2018, Cain2025}. High-redshift galaxy surveys, including from the \textit{James Webb Space Telescope} (\text{JWST}), directly probe the source population and have reported significant field-to-field variation, consistent with a patchy, inhomogeneous reionization scenario with substantial large-scale dispersion around a volume-averaged neutral fraction \citep{Kashino2023, Nakane2024, Napolitano2024, Kakiichi2025}. Individual sightlines and survey fields sample particular environments, whereas sky-integrated signals probe a much broader average, and galaxies provide information on population statistics. Combining these self-consistently is a non-trivial task and subject to astrophysical and systematic uncertainties \citep{Naidu2020, Matthee2022, Kageura2025, Shimizu2025}.

Additional CMB anisotropy probes are also sensitive to the reionization morphology, e.g.\ the kinetic Sunyaev--Zel'dovich (kSZ) signal further constrains the duration and topology of reionization \citep{Zahn2012, Calabrese2014, SmithFerraro2017, Paul2021, Nikolic2023, Chen2023_PatchykSZ}. More generally, patchy screening and related cross-correlation formalisms connect inhomogeneous reionization to polarization and temperature anisotropies \citep{Dvorkin2009Bmode, Battaglia2013, Gluscevic2013, Natarajan2013, Kramer2025}, including the potential role of helium ionization patchiness \citep{Caliskan2024}. The fluctuations can also depend on baryonic physics beyond radiative transfer sources and sinks \citep{Park2018}. Reionization constraints are also crucial for interpreting CMB polarization measurements used to constrain primordial gravitational waves \citep{Jiang2025}.

Interpreting these data requires care because $\tau_\text{CMB}$ alone does not select a unique reionization history. There are degeneracies among the midpoint, duration, and overall shape of the ionization history \citep{Doussot2019}. This has motivated flexible parametrizations that permit asymmetric histories \citep{Trac2018}, and some analyses argue that simple symmetric histories are disfavored by current data \citep{Paoletti2025}. In parallel, independent measurements that avoid or downweight low-$\ell$ CMB polarization data have obtained somewhat higher optical depths \citep{Giare2024, Kageura2026}, while related DESI LRG--CMB-lensing measurements provide complementary lower-redshift growth information relevant to these parameter comparisons \citep{Sailer2025}. More directly, recent studies have examined how cosmological parameters shift when $\tau_\text{CMB}$ is relaxed relative to the Planck low-$\ell$ polarization prior \citep{Elbers2025, Sailer2026}. In the DESI BAO context, \citet{Sailer2026} showed that adopting a larger optical depth can reduce the BAO--CMB mismatch in $\Omega_m$ and change extended-parameter inferences, including those involving evolving dark energy and the summed neutrino mass. This is a conditional cosmological inference rather than an astrophysical determination of the reionization history: a value substantially above the Planck low-$\ell$ optical depth would still need to be produced by additional high-redshift ionization, for example through evolving escape fractions \citep[e.g.][]{Yeh2023}, early luminous source populations, or additional ionizing channels. Any such interpretation benefits from a clear accounting of how patchiness and averaging choices enter $\tau_\text{CMB}$ predictions.

A subtle but important point is that $\tau_\text{CMB}$ is intrinsically an electron-weighted integration along a light cone. In an inhomogeneous IGM, the first ionized regions preferentially reside in overdense environments with earlier structure formation. The resulting density--ionization correlation implies that the optical depth is more closely matched by using the mass-weighted global ionized fraction in the homogeneous Universe expression than by treating the ionized volume filling fraction as the relevant average. Indeed, some previous works have explicitly adopted mass-weighted histories in optical-depth calculations \citep{Trac2018, Trac2022, Glazer2018}, while some observationally motivated parameterizations and simulation analyses are phrased in terms of volume-averaged ionized or neutral fractions \citep[e.g.][]{Mesinger2011, Planck2016, Greig2017, Naidu2020, Kannan2022a}. This distinction is important to state explicitly, even though it naturally emerges from mock observations. Specifically, reionization patchiness implies that $\tau_\text{CMB}$ is itself a spatial field with sightline-to-sightline fluctuations \citep[e.g.][]{dvorkin2009reconstructing, Caliskan2024}, motivating direct, simulation-based estimates of both the mean optical depth and its variance, in particular to avoid inconsistencies in modeling assumptions and ambiguity between mass-weighted and volume-weighted ionization fractions.

Quantifying these higher-order effects self-consistently is challenging. It requires large cosmological volumes to capture reionization variance on $\gtrsim 100$\,cMpc scales \citep{Iliev2014}, and even larger volumes to robustly characterize large-scale 21\,cm statistics \citep[$\gtrsim 250$\,cMpc;][]{kaur2020minimum}, together with sufficient resolution and radiative transfer to follow the topology of ionization fronts and its correlation with the evolving density field \citep{Neyer2024, Jamieson2025}. This paper presents early results from the new \lumina project, which was designed adjacent to the \thesan project \citep{Kannan2022a, Smith2022, Garaldi2022, Garaldi2024} to simultaneously address several of these requirements using a much larger-scale ($500$\,cMpc) galaxy-driven radiation-hydrodynamical simulation with on-the-fly radiative transfer and high-cadence spatial and light-cone outputs \citep{Zier2026}. In this paper, we use these outputs to examine the CMB optical depth as a field, including analysis of the mean, sightline variance and non-Gaussianity, and angular statistics, comparable to other observables such as 21\,cm tomography \citep{Mellema2006}. In particular, we: (\textit{i}) quantify the light-cone boost that arises from sampling many overdense, early-ionized regions along long sightlines, (\textit{ii}) show how and why mass-weighted ionization histories reproduce this boost, (\textit{iii}) measure the scale dependence of the optical-depth variance under coarse-graining, and (\textit{iv}) present angular power spectra and real-space correlation functions for the light cone generated $\tau_\text{LOS}$ decomposed into \HII, \HeII, and \HeIII contributions, including their cross-correlations.

The paper is organized as follows. In Section~\ref{sec:methods}, we summarize the \lumina simulation and the Cartesian and light-cone outputs used to compute $\tau_\text{CMB}$. In Section~\ref{sec:results}, we present maps, mean histories and boost factors, fluctuation statistics, and angular/real-space correlation measures. In Section~\ref{sec:discussion}, we discuss implications for computing $\tau_\text{CMB}$ from models and simulations and for interpreting current optical-depth constraints. We summarize our main conclusions in Section~\ref{sec:conclusions}. Appendices~\ref{ap:retiling}--\ref{ap:divergence} document the light cone retiling choice, resolution tests, and mass- versus volume-weighted ionization differences.

\section{Methods}
\label{sec:methods}

\subsection{Simulation overview}

The \lumina simulation is a large-volume cosmological radiation-hydrodynamics (RHD) calculation designed to follow the coupled evolution of galaxies, black holes, and the IGM through hydrogen and helium reionization down to $z=3$ \citep{Zier2026}. \lumina evolves a comoving cube of side length $L_\text{box} = 500\,\text{cMpc}$ with $2 \times 6000^3$ resolution elements (gas cells and dark matter particles), corresponding to baryonic and dark-matter mass resolutions of $m_\text{gas} = 3.6\times10^6\,\Msun$ and $m_\text{DM} = 1.9\times10^7\,\Msun$. Gravitational forces employ Plummer-equivalent softening lengths of $\epsilon_\text{gas,min} = 0.44\,\text{ckpc}$ (minimum, for gas) and $\epsilon_\text{DM} = 1.77\,\mathrm{ckpc}$ (fixed, for dark matter).

\lumina is run with the moving-mesh code \arepo \citep{springel2010pur, pakmor2016improving, weinberger2020arepo} and employs the radiative-transfer module \areport \citep{Kannan2019}, which solves the radiative-transfer moment equations with an M1 closure \citep{Levermore1984, Dubroca1999}. The overall code base incorporates scalability improvements developed for large-volume simulations \citep{Gadget4, pakmor2023millenniumtng} and GPU acceleration \citep{zier2024adapting}. For context, the complementary \thesanone simulation \citep{Kannan2022a, Smith2022, Garaldi2022, Garaldi2024} provides fully-coupled RHD in a smaller volume, while MillenniumTNG \citep[MTNG;][]{pakmor2023millenniumtng} provides a larger volume without on-the-fly radiative transfer but run down to $z = 0$. Specifically, \thesanone and MTNG have gas resolutions of $5.8\times10^5\,\Msun$ and $3.1\times10^7\,\Msun$ throughout $95.5$\,cMpc and $740$\,cMpc volumes for $2100^3$ and $4320^3$ particles, respectively. The adopted cosmological parameters are those of the Planck 2018 baseline TT,TE,EE+lowE+lensing+BAO constraints under a $\Lambda$CDM model ($\Omega_m = 0.3096$, $\Omega_b = 0.04897$, $H_0 = 67.66\,\text{km}\,\text{s}^{-1}\,\text{Mpc}^{-1}$, $\sigma_8 = 0.8102$, $n_s = 0.9665$) and the initial redshift is $z_\text{init} = 49$.

\subsection{Hydrodynamics, self-gravity, and galaxy formation}

\arepo solves the Euler equations with a second-order finite-volume Godunov scheme on an unstructured, moving Voronoi mesh \citep{springel2010pur, pakmor2016improving}. The quasi-Lagrangian mesh motion reduces advection errors relative to static grids, preserves Galilean invariance, and allows natural adaptive spatial resolution, which is advantageous for cosmological flows with large bulk velocities \citep{vogelsberger2012moving}. Time integration utilizes a hierarchical time-step structure with second-order accuracy and gas cells are refined and derefined to maintain an approximately constant target mass. Self-gravity between gas, dark matter, stars, and black holes is computed with an adaptive Tree--PM approach, combining a particle--mesh method for long-range forces and a hierarchical tree for short-range forces \citep{Bagla2002, Bagla2003, Barnes1986, Aarseth2003}.

Galaxy formation follows the IllustrisTNG subgrid model \citep{Vogelsberger2013, IllustrisIntro, IllustrisNature, Weinberger2017, Pillepich2018Model}. In brief, star formation proceeds stochastically with a Kennicutt--Schmidt-motivated density dependence in cold, dense gas ($n_\text{H} > 0.106\,\text{cm}^{-3}$) treated as an unresolved two-phase interstellar medium equation of state \citep[EoS;][]{Springel2003}. Stellar feedback is implemented through an effective galactic-wind model, in which hydrodynamically decoupled kinetic winds are launched from star-forming cells with prescribed velocity, energy loading, and metal loading, together with enrichment from stellar evolution. Black holes are seeded in sufficiently massive haloes (above $\sim 5\times10^{10}\,h^{-1}\,\Msun$), grow via Eddington-limited Bondi--Hoyle accretion and mergers, and provide feedback in both thermal (high-accretion quasar) and kinetic (low-accretion inefficient wind) modes \citep{Weinberger2017, Pillepich2018Model, Bulichi2025}. These prescriptions are adopted in \lumina with minor modifications that accommodate the computational requirements of a $500$\,cMpc RHD run \citep{Zier2026}. In particular, \lumina tracks total metallicity rather than individual elemental abundances, deposits newly produced metals to the nearest gas cell after $z < 4.75$ to reduce neighbor-search cost, and updates the coupling between the effective ISM, radiative transfer, and feedback relative to earlier \thesan implementations. Overall, the simulation self-consistently links reionization to the evolving population of galaxies and active galactic nuclei (AGN). At higher ISM and CGM resolution, the complementary \thesanzoom suite explores how the external reionization environment and multi-phase ISM physics shape early galaxies \citep{Kannan2025Zoom, Zier2025Zoom}.

\subsection{Radiation transport and thermochemistry}

\lumina evolves ionizing radiation on the same moving mesh using a moment-based radiative-transfer solver with the M1 closure \citep{Levermore1984, Dubroca1999} implemented in \areport\ \citep{Kannan2019, zier2024adapting}. Photoionization and photoheating are coupled to a non-equilibrium primordial thermochemistry network that tracks the ionization states of hydrogen and helium and the associated cooling and heating processes. To reduce the radiative-transfer time-step constraint, the solver employs a reduced-speed-of-light approximation \citep{Gnedin2001} with a value of $\tilde{c} = 0.2\,c$, which has been shown to yield converged reionization histories for similar setups \citep{Kannan2022a}. The radiative-transfer and chemistry are subcycled within each hydrodynamical time step (64/256 times before/after optimizing for helium reionization at $z = 4.75$), all performed with GPU acceleration for efficiency.

During hydrogen reionization ($z>4.75$), the radiation field is discretized into six energy bins spanning 13.6\,eV to 2\,keV. The dominant ionizing sources are stellar populations, with emissivities based on BPASS binary population synthesis models \citep[v2.2.1;][]{BPASS2017, Stanway2018} and a Chabrier initial mass function \citep{Chabrier2003}. Accreting black holes provide harder spectra that are essential for helium reionization \citep{mcquinn2009he}, for which \lumina adopts a composite AGN spectral energy distribution \citep{Shen2020}. \lumina additionally includes high-energy X-ray components motivated by high-mass X-ray binaries and hot interstellar gas \citep{2013Fragos-HMXB, 2016Fragos-Erratum, Madau2017, 2012bMineo-HotISM, 2014Pacucci}, which can pre-heat the IGM \citep{Pritchard2007, ma2018, 2018Eide, 2020Eide}. At $z=4.75$, the simulation transitions to a mode optimized for \HeII$\rightarrow$\HeIII reionization. In this mode, all photons with energies below the \HeII edge (54.42\,eV) are replaced by a spatially uniform metagalactic background, while photons at higher energies are merged to reduce computational cost. AGN are retained as the self-consistent radiation sources that drive \HeIII bubbles. This split is convenient for $\tau_\text{CMB}$ because hydrogen reionization dominates the optical-depth budget, while later helium reionization provides a smaller but spatially structured contribution.

\subsection{Cosmological initial conditions}

Initial conditions are generated at $z_\text{init} = 49$ using a linear matter power spectrum computed with separate transfer functions for baryons and dark matter, and including the baryon--dark matter streaming velocity \citep{Zier2026}. The perturbations are imprinted using second-order Lagrangian perturbation theory onto a two-component glass configuration made up of particles representing dark matter and gas, each in gravitational equilibrium prior to imprinting the perturbations. The density perturbations themselves are realized through spatial displacements to account for the mass-weighted density fluctuations of dark matter and baryons, and by particle mass perturbations \citep{Hahn:2021aa} to account for the difference in dark matter and gas density fields. This approach ensures that small-scale power is retained in both components, and that the subsequent evolution of the baryon and CDM fluctuations reflects the transfer-function differences without washing them out by numerical discreteness effects. The simulation adopts the Planck 2018 $\Lambda$CDM cosmological parameters, for which the combined temperature and polarization analysis inferred $\tau_\text{CMB} = 0.0544\pm0.0073$ \citep{Planck2020}. For comparison, re-analyses of the Planck polarization data update the inferred optical-depth likelihood, rather than the cosmological parameters used in our initial conditions, reporting $\tau = 0.0566^{+0.0053}_{-0.0062}$ (polarization only) and $\tau = 0.059\pm0.006$ when combined with temperature data \citep{pagano2020reionization}.

\begin{figure*}
    \centering
    \safeincludegraphics[width=\textwidth]{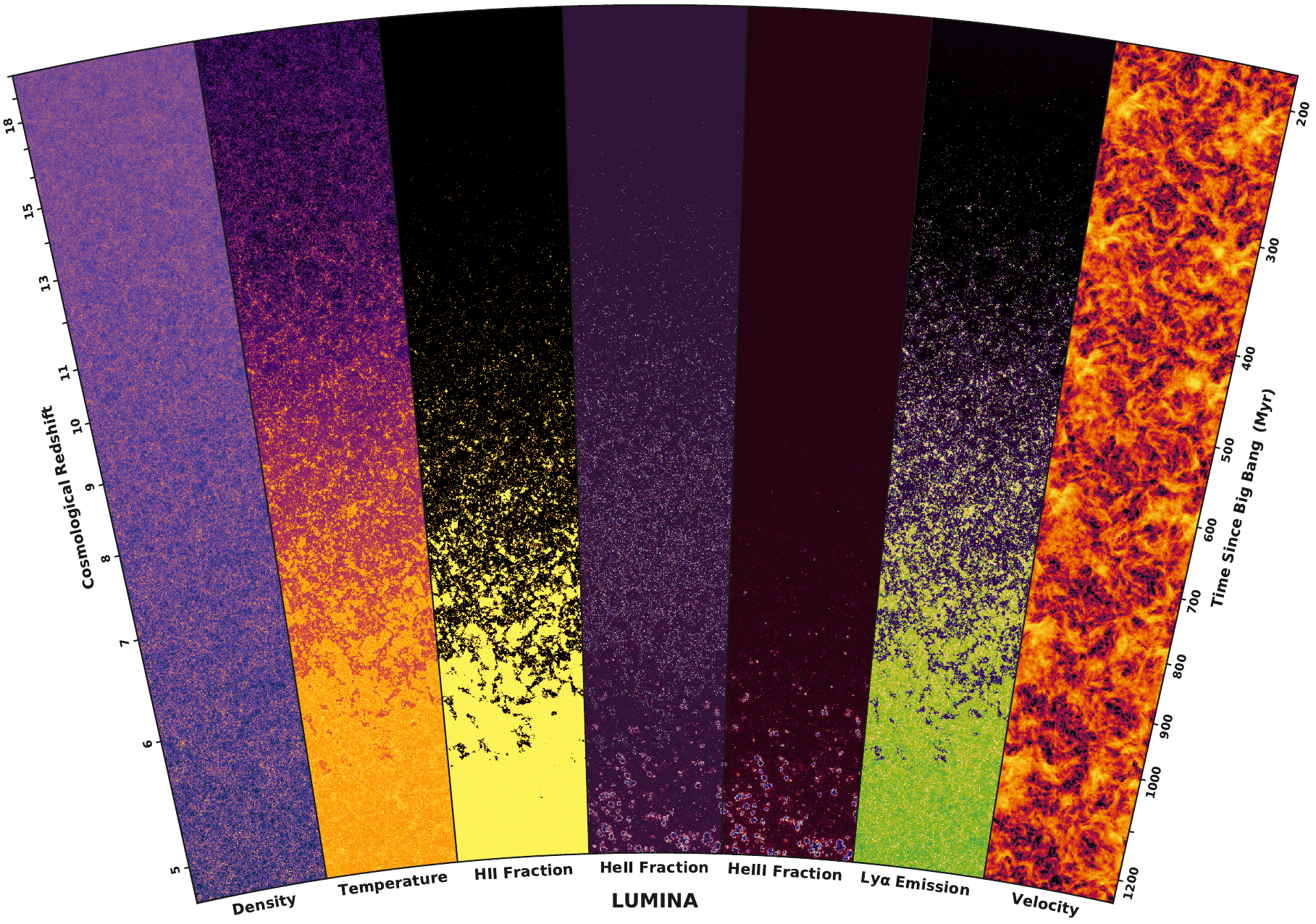}
    \caption{Composite light-cone fan through the \lumina simulation, showing line-of-sight columns for total gas density, temperature, \HII, \HeII, and \HeIII fractions, Ly$\alpha$ emissivity, and velocity amplitude (norm), over a $3.6^\circ$ opening angle across the EoR redshift range ($z \geq 4.75$). The box has a side length of 500\,cMpc and $6000^3$ resolution elements, with on-the-fly adaptive cloud-in-cell deposition onto a $5120^2$ light-cone grid. Color maps emphasize the patchy distribution of ionized bubbles, their temperature patterns, and the filamentary cosmic web.}
    \label{fig:lightcone_fan}
\end{figure*}

\subsection{High-cadence Cartesian outputs and light cones}

\lumina produces two complementary analysis products that allow us to precisely characterize the evolving IGM (Cartesian grids) and reconstruct realistic electron column densities (light cones).

First, the simulation outputs high-cadence Cartesian grids at intervals of $\simeq 2$--$4$\,Myr. Mass- and volume-weighted IGM fields (e.g.\ gas density, temperature, velocity, ion fractions, and radiation energy) are deposited onto a $1280^3$ uniform grid (with a fixed cell size of $390\,\text{ckpc}$) using an adaptive, second-order cloud-in-cell procedure in which each Voronoi cell is approximated as a cube and its mass and/or volume is distributed to neighboring voxels in proportion to geometric overlap. These outputs capture the three-dimensional structure of ionization fronts and are used in this paper to explore the spatial variation of quantities throughout the box.

Second, \lumina constructs a small-angle light cone on the fly, designed to sample the rapidly evolving reionization topology without snapshot-to-snapshot interpolation. This comprises $5120^2$ rays anchored at the box center. The transverse pixel size is chosen so that the cone spans the full $500$\,cMpc simulation volume at $z=5$, corresponding to a field of view of $3.6^\circ$; for $z>5$ the cone is filled by periodic replications of the box, extending to $z=30$. Along each ray, the line of sight is subdivided into voxels whose depth matches the transverse pixel size, yielding approximately cubic sampling elements. At every global time step, the code deposits gas properties from intersected Voronoi cells (again approximated as adaptive cloud-in-cell cubes) into the corresponding voxels. Both real-space and redshift-space versions of the cone are stored. In post-processing, we apply rotation and reflection symmetries to reduce periodic repetition artifacts, and we quantify the effect of this retiling strategy in Appendix~\ref{ap:retiling}, concluding that tiling 75\% of the box minimizes variance in integrated light-cone maps. Each ray contains 47,169 voxels between $z=30$ and $z=3$. For convergence tests in Appendix~\ref{ap:resolution}, we also analyze coarsened versions of the cone constructed by factors-of-two reductions (preserving proper weightings) in the transverse and radial directions down to $5^2$ resolution.

\begin{figure*}
    \centering
    \safeincludegraphics[width=\textwidth]{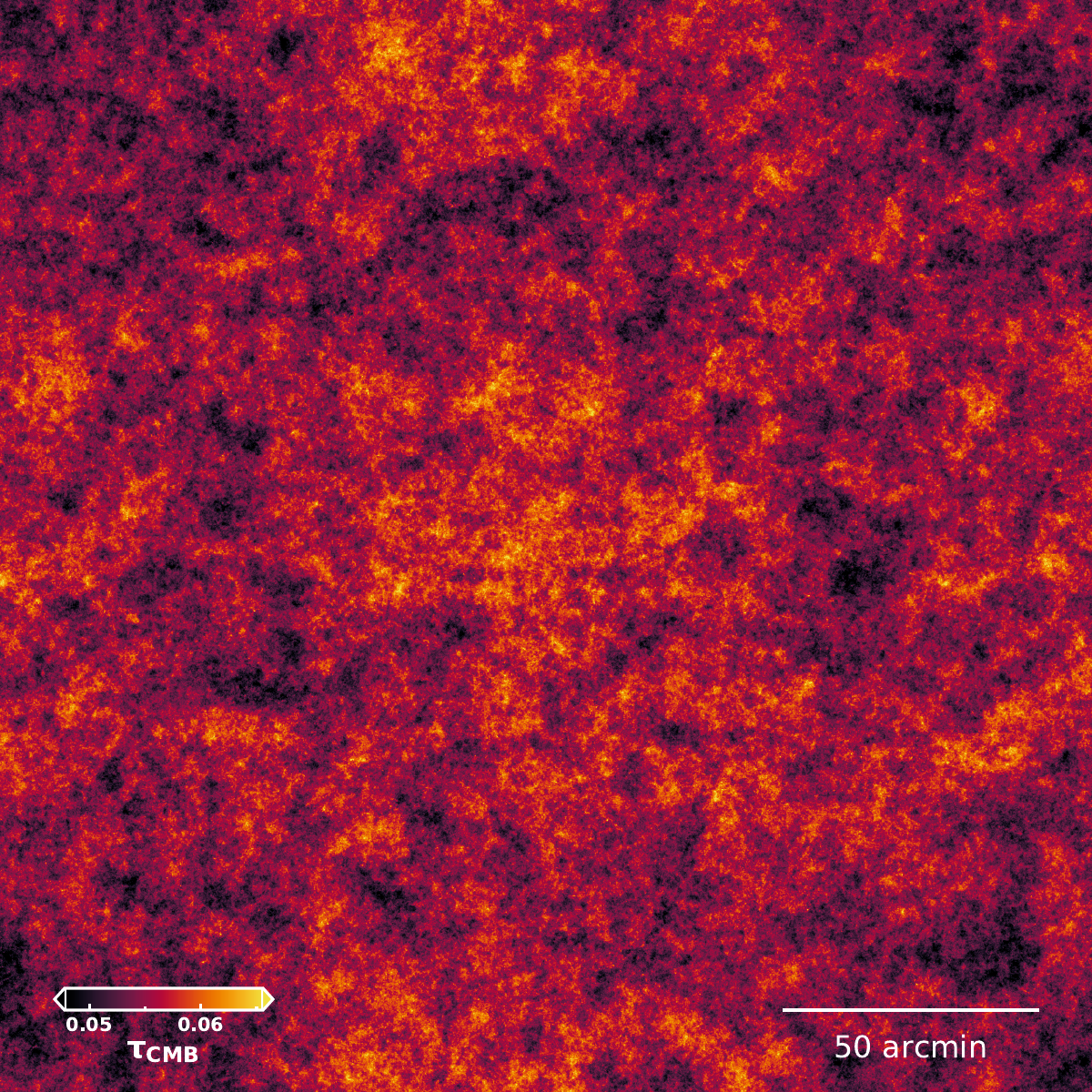}
    \caption{Spatial image of the CMB optical depth $\tau_\text{CMB}$ integrated along each sightline across the full $3.6^\circ$ light-cone field of view. The map is constructed by assembling rotated and transposed tiles covering 75\% stretches of the simulation volume to minimize periodic repetition. Although this reduces the overall variance between sightlines, the map retains substantial coherent fluctuations, with the standard deviation $\approx 5.0$--$6.5\%$ of the global average depending on the retiling choice and extrema reaching up to $\pm50$\%.}
    \label{fig:lightcone_image}
\end{figure*}

\subsection{CMB optical depth}

The Thomson optical depth out to redshift $z$ is
\begin{equation}
    \tau_\text{CMB}(<z)
    = c\,\sigma_\text{T}\int_{0}^{z} \frac{n_e(z')}{(1+z')\,H(z')}\,\text{d}z' \, ,
    \label{eq:opticalDepthCMB}
\end{equation}
where $c$ is the speed of light, $\sigma_\text{T}$ is the Thomson cross section, $n_e$ is the proper free-electron number density, and $H(z)$ is the Hubble parameter. In the simulation,
\begin{equation}
    n_e = n_\text{H}\,x_\HII + n_\text{He}\,\left(x_\HeII + 2\,x_\HeIII\right) \, ,
\end{equation}
with $n_\text{H}$ and $n_\text{He}$ obtained from the gas density assuming a primordial composition with helium mass fraction $Y_\text{p}=0.24$. The helium terms count one free electron from \HeII and two from \HeIII. The averaging distinction can be made explicit by writing the density-weighted contribution for any ionization variable $x$ as \citep{Trac2018}
\begin{equation} \label{eq:density_weighting}
  \langle x \rho \rangle_V = \frac{\int x\,\rho\,\text{d}V}{\int \text{d}V} = \frac{\int x\,\rho\,\text{d}V}{\int \rho\,\text{d}V} \frac{\int \rho\,\text{d}V}{\int \text{d}V} = \bar{\rho} \, \langle x \rangle_m \, ,
\end{equation}
where $\bar{\rho} \equiv \langle \rho \rangle_V$ is the volume-weighted mean density. A homogeneous calculation based on the volume-weighted ionized fraction instead approximates this term as $\bar{\rho} \langle x \rangle_V$, dropping the density--ionization correlation that is present during inside-out reionization.

We compute several versions of $\tau_\text{CMB}$ for $z\ge z_\text{end}=3$: (\textit{i}) global optical depth constructed from high-cadence volume-weighted and mass-weighted mean ion fractions, (\textit{ii}) a pixelwise light-cone optical-depth map obtained by discretizing Eq.~(\ref{eq:opticalDepthCMB}) to perform spatially-resolved line-of-sight (LOS) integrations along each ray based on volume-weighted ion fractions within each voxel, yielding both a map and a sightline-averaged value $\langle \tau_\text{LOS} \rangle$, and (\textit{iii}) co-spatial integrated optical depths obtained by selecting subvolumes and integrating their local volume-weighted ionization histories. In all cases, for $z < z_\text{end}$ we assume hydrogen and helium are fully ionized (\HII and \HeIII), implying $n_e / n_\text{H} = 1 + 2\,n_\text{He}/n_\text{H} \simeq 1.158$ for a primordial helium mass fraction $Y_\text{p} = 0.24$. This homogeneous low-redshift contribution is included in all total optical depths and gives $\tau_\text{CMB}(z < 3) = 0.0152$, or about $28\%$ of the fiducial light-cone mean. It is beyond the scope of this work to model the residual spatial fluctuations from the low-redshift cosmic web.

\begin{figure*}
    \centering
    \safeincludegraphics[width=\textwidth]{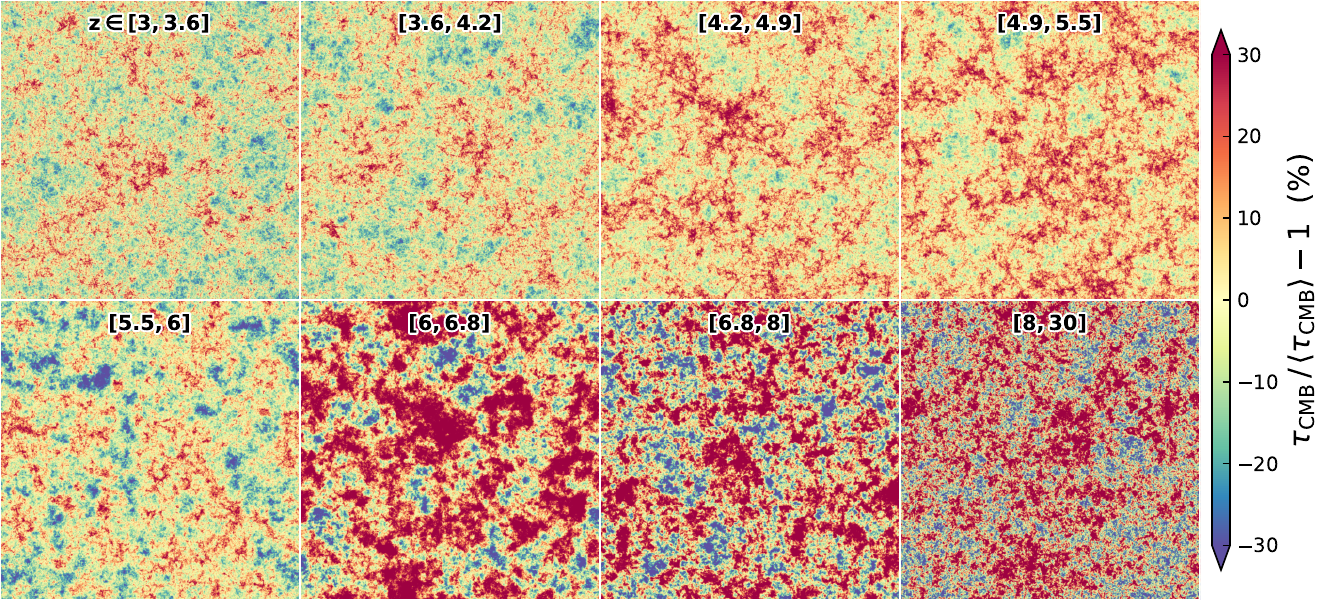}
    \caption{Spatial image of the CMB optical depth $\tau_\text{CMB}$ divided into eight segments illustrating the contribution of successive redshift intervals to the light cone. Each panel shows a range corresponding to one eighth of the total optical depth accumulated over $z > 3$. Color scales are normalized independently to highlight relative fluctuations within each panel. The changing morphology reflects the evolving reionization topology from small bubbles above $z \gtrsim 8$ to large ones by $z \sim 6$--$7$.}
    \label{fig:lightcone_image_ranges}
\end{figure*}

\section{Results}
\label{sec:results}

\subsection{Morphology and maps of $\tau_{\rm CMB}$}

We begin by visualizing the evolving reionization topology and the resulting optical depth contributions in \lumina to set the stage for later quantitative analysis. In Fig.~\ref{fig:lightcone_fan}, we show a composite light-cone history tracing a mock observation through the simulation. The wide $3.6^\circ$ slice spans the 500\,cMpc box at $z = 5$, yet the simulation resolves individual ionized bubbles down to galaxy scales. The density and temperature maps illustrate the familiar cosmic-web morphology in which high-density filaments and clusters are surrounded by cool underdense voids. The ionization fields show that reionization proceeds from the densest regions outward, \HII regions (third row) percolate first along the same large-scale structures, while \HeIII (fifth row) lags until the onset of harder radiation sourced by AGN. The diffuse Ly$\alpha$ emissivity ($\propto \rho^2$) traces regions of ionizing-photon production, thereby revealing the interplay between small-scale features and large-scale topology. The velocity map highlights coherent flows into overdense nodes, also showing the periodic structure of the box that is less evident in other images.

In this context, we view patchy reionization as the spatial and temporal patterns that CMB photons traverse through. Early sources cluster in overdensities, creating high electron column densities and an optical-depth bias relative to a uniform Universe. The light-cone geometry ensures that each sightline encounters different structures at different redshifts, with the rapid sweeps effectively sampling the global topology. The spatially-correlated enhancement in ionization with density explains why the light-cone $\tau_\text{LOS}$ distribution contains a range of integrated values rather than being described by a single representative ray.

\begin{figure*}
    \centering
    \safeincludegraphics[width=\textwidth]{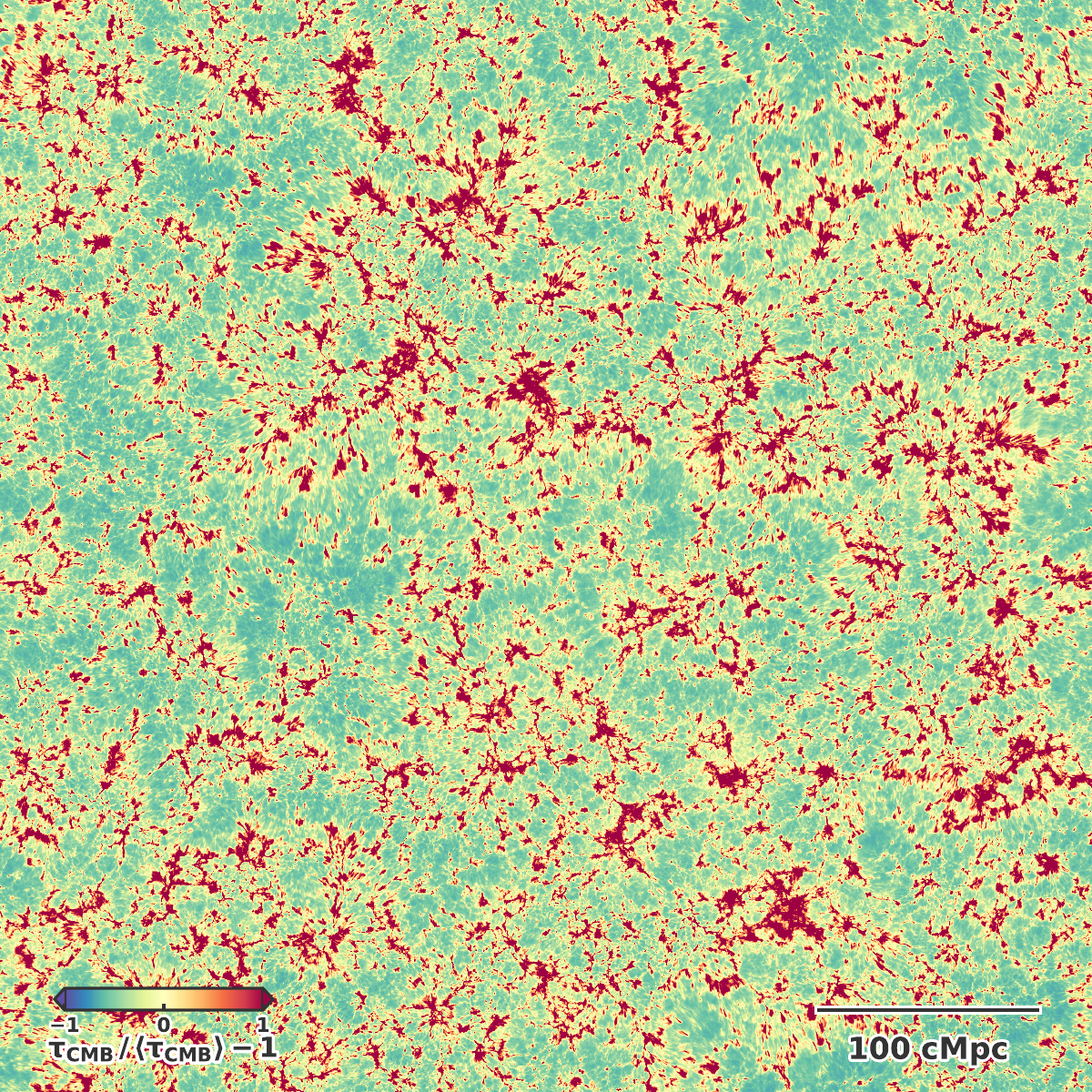}
    \caption{Spatial image of the CMB optical depth $\tau_\text{CMB}$ obtained by integrating through the local reionization history of fixed co-spatial regions rather than through an evolving light cone, yielding a maximal-variance estimate. This is achieved by integrating the local reionization history in the same overdense or underdense regions. The image is a slice from the 3D volume, with a resolution of 1280 pixels across or $390\,\text{ckpc}$ per pixel, showing the cosmic web with persistently ionized high-density structures. Colors indicate the relative optical depth enhancement compared to the global average. For clarity, no light-cone effects are included.}
    \label{fig:local_image}
\end{figure*}

The integrated optical depth collapses the redshift axis in the light cone to a single observed value per sightline. In Fig.~\ref{fig:lightcone_image}, we show the spatial image of $\tau_\text{LOS}$ for each pixel accumulated across all cosmic time. The simulated volume captures coherent features with broad large-scale smoothness modulated by smaller-scale granular fluctuations. The variation is caused by peaks corresponding to overdense regions where ionized bubbles first grow and overlap, while troughs arise from extended voids or late-ionizing patches. As mentioned, the map is constructed using a tiling strategy, with symmetry transformations applied every 377\,cMpc, designed to suppress periodic resampling of nearly aligned structures and minimize variance across pixels (see Appendix~\ref{ap:retiling}).

Quantitatively, and further summarized in Table~\ref{tab:optical_depth_stats}, the average value across all sightlines is $\langle \tau_\text{LOS} \rangle = 0.055$, which we emphasize is independent of the detailed structure of the image; i.e.~while retiling alters the apparent morphology it does not change the mean. We consider the standard deviation $\sigma_\text{LOS} = 0.00277$ (or $0.00359$ without retiling) to be a conservative lower (upper) limit, which is approximately $5.0\%$ ($6.5\%$) of the average. The distribution spans a wide range of $\tau_\text{LOS}$ values, with the minimum ($0.044$) and maximum ($0.097$) sightlines below and above the mean by $20\%$ and $75\%$, respectively. The spatial patterns appear to be coherent on degree scales, reflecting the characteristic size of the largest (and rarest) ionized structures emerging by $z \approx 6$--$8$ of up to $\sim 100\,\text{cMpc}$.

\begin{table}[t]
\centering
\addtolength{\tabcolsep}{3pt}
\renewcommand{\arraystretch}{1.1}
\begin{tabular}{lcccc}
\toprule
Statistic & Total & \HII & \HeII & \HeIII \\
\midrule
\multicolumn{5}{l}{Global mass-weighted average} \vspace{0.1cm}\\
$\tau_{\text{CMB},m}$ & 0.0544 & 0.0489 & 0.00223 & 0.00329 \\
\midrule
\multicolumn{5}{l}{Global volume-weighted average} \vspace{0.1cm}\\
$\tau_{\text{CMB},V}$ & 0.0515 & 0.0462 & 0.00203 & 0.00329 \\
\midrule
\multicolumn{5}{l}{Light-cone (LOS) mean} \vspace{0.1cm}\\
$\langle \tau_\text{LOS}\rangle$ & 0.0550 & 0.0494 & 0.00221 & 0.00342 \\
\midrule
\multicolumn{5}{l}{Light-cone (LOS) median} \vspace{0.1cm}\\
$\tilde{\tau}_\text{LOS}$ & 0.0549 & 0.0493 & 0.00220 & 0.00340 \\
\midrule
\multicolumn{5}{l}{Light-cone (LOS) standard deviation} \vspace{0.1cm}\\
$\sigma_\text{LOS}$ & 0.00277 & 0.00252 & 0.000161 & 0.000182 \\
\bottomrule
\end{tabular}
\caption{\justifying \noindent \textup{Summary of optical depth statistics for the \lumina light cone. Global mass-weighted ($\tau_{\text{CMB},m}$) and volume-weighted ($\tau_{\text{CMB},V}$) averages are over the full 500\,cMpc box. The light-cone mean ($\langle \tau_\text{LOS}\rangle$) and median ($\tilde{\tau}_\text{LOS}$) provide line-of-sight (LOS) statistics across all pixels, while the standard deviations quantify the pixel-to-pixel scatter in the light-cone maps. Columns list the total optical depth and the contributions from \HII, \HeII, and \HeIII, including the simulated $3 \le z \le 30$ contribution plus the homogeneous fully ionized $z<3$ contribution. Comparing the mass-weighted and volume-weighted rows shows the enhancement from density--ionization correlations. The He contributions remain small compared with \HII for all statistics.}}
\label{tab:optical_depth_stats}
\addtolength{\tabcolsep}{-3pt}
\renewcommand{\arraystretch}{1}
\end{table}

The morphological evolution of the contribution to $\tau_\text{LOS}$ is illustrated in Fig.~\ref{fig:lightcone_image_ranges}, which partitions the light-cone integral into eight equal optical-depth segments across redshift. The highest redshift panel ($z = 8$--$30$) shows scattered, compact ionized regions embedded in a mostly neutral Universe. At these early times the variance in the projected $\tau_\text{LOS}$ is dominated by rare high-density ionized regions and extended neutral regions. As redshift decreases ($z = 6.8$--$8$), the \HII regions expand and overlap in a competition of structured growth, producing coherent fluctuations on a range of angular scales. During the large-scale coalescence of dominant bubbles ($z = 6$--$6.8$), the structure has evolved into discernible regions matching the characteristic bubble size distribution. Finally, at lower redshifts ($z = 3$--$6$), the maps become smoother as the IGM approaches full ionization and residual contributions increasingly trace the large-scale structure of the Universe.

For comparison, in Fig.~\ref{fig:local_image}, we also visualize a representative slice of the three-dimensional volume in which we integrate through the local reionization history of each region. By ignoring light-cone traversal effects and remaining fixed on a single position we isolate the spatial contribution to $\tau_\text{CMB}$ for an upper-limit style variance estimate of repeatedly sampling the same regions. Thus, the contrast of large-scale structure is amplified as overdense regions ionized at earlier epochs exhibit high static optical depths, while underdense voids are comparatively low. However, the exaggerated fluctuations deliberately demonstrate the clumping bias induced by the early ionization boost in these dense environments. In reality, the effective optical depth along a light cone blends epochs and uncorrelated structures, and thus partially smooths these local fluctuations. Still, the static co-spatial map helps to clarify the physical origin of the effect.

This figure also suggests that if we average over sufficiently large volumes the fluctuations in $\tau_\text{CMB}$ will be suppressed until it converges to the (lower) global volume-weighted average. However, dividing into even smaller subvolumes resolves more of the $\rho$--$x_\HII$ correlated clumping, and the mean climbs until the majority of ionized bubbles are resolved so sightlines intersect high-density ionized regions instead of mixed-density and partial-ionization gas. Smoothing on a given scale erases the clumpy, multi-phase topology of reionization. As a result, calculations without resolved substructure can infer different ionization states depending on whether they use volume- or mass-weighted quantities, illustrating why the chosen aperture matters.

\begin{figure}
    \centering
    \safeincludegraphics[width=\columnwidth]{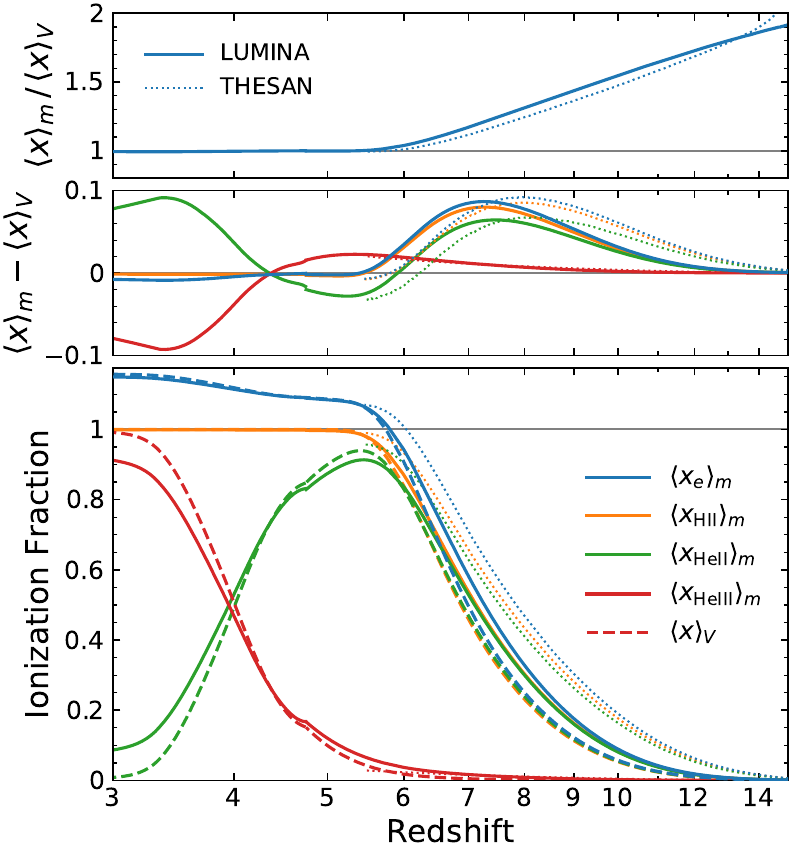}
    \caption{\textit{Top panel:} Ratio of the mass-weighted to volume-weighted electron fraction, which produces a $\langle x \rangle_m / \langle x \rangle_V = \max\{1, 1.033\,\ln(z) - 0.835\}$ boost factor that propagates into light-cone calculations. \textit{Middle panel:} Differences between the mass- and volume-weighted fractions, $\langle x \rangle_m - \langle x \rangle_V$, showing the bias from the patchy bubble structure. \textit{Bottom panel:} Redshift evolution of the global mass-weighted (solid curves) and volume-weighted (dashed curves) ionization fractions for electrons, \HII, \HeII, and \HeIII, illustrating the progress of hydrogen and helium reionization. For reference, we include the same quantities from the \thesanone simulation \citep[dotted curves;][]{Kannan2022a}, which shows that the same qualitative behavior appears even with an earlier reionization history (to avoid overcrowding only mass-weighted fractions are shown for \thesanone in the bottom panel).}
    \label{fig:x_boost}
\end{figure}

\subsection{Light cones probe the mass-weighted history}

The CMB optical depth depends not only on the timing of reionization but also on how the free-electron density is averaged. A common practice is to compute $\tau_\text{CMB}$ by inserting a global ionization history into Eq.~(\ref{eq:opticalDepthCMB}). In a patchy IGM, however, the expectation value of the electron density involves $\langle n_e \rangle$, so correlations between overdensity and ionization state matter. For additional context on the volume- and mass-weighted differences, Fig.~\ref{fig:x_boost} shows the redshift evolution of the ionization fractions entering into the global $\tau_\text{CMB}$ calculations and the corresponding mass-weighting boost.

In the top panel, we show the ratio of the mass-weighted to volume-weighted electron fraction, which provides a boost in $\tau_{\text{CMB},m} / \tau_{\text{CMB},V}$. This ratio is straightforwardly motivated by considering the overdensity ($\Delta \equiv \rho / \bar{\rho}$) bias of ionized regions:
\begin{equation}
  \langle \Delta \rangle_x \equiv \frac{\int \Delta\,x \,\text{d}V}{\int x\,\text{d}V} = \frac{\int x\,\rho \,\text{d}V}{\int \rho\,\text{d}V} \frac{\int \text{d}V}{\int x\,\text{d}V} = \frac{\langle x \rangle_m}{\langle x \rangle_V} \, .
\end{equation}
This factor is exactly the boost one would expect by traversing through overdense ionized regions, which can also be understood in the following conversion from total to ionized hydrogen number density in Eq.~(\ref{eq:density_weighting}). We note however that the light-cone values can still be different due to local variations from the global mean encountered while sweeping through the box. Moreover, spatial fluctuations are inaccessible without ray-based integrations. Higher order statistics like $\langle x^2 \rangle$ may also capture the magnitude of the variance, although we do not explore that possibility in this paper. For use in semi-numerical frameworks, we fit the boost factor as $\langle x \rangle_m / \langle x \rangle_V = \max\{1, 1.033\,\ln(z) - 0.835\}$.

The middle panel emphasizes the difference between the mass- and volume-weighted fractions, $\langle x \rangle_m - \langle x \rangle_V$. The \HII values can lead by up to $10\%$, while the helium ones have slightly more complex deviation histories. The bottom panel shows the redshift evolution of the ionization fractions themselves. The \HII fraction follows a late-reionization history, rising from near zero at $z \sim 15$ to a midpoint around $z \sim 7$ and completion by $z \sim 5$. \HeII closely mirrors the \HII evolution but peaks below $95\%$ as \HeIII regions begin to emerge. The \HeIII fraction then grows more gradually, with a midpoint at $z \sim 4$ and volume-weighted completion at $z \sim 3$ with residual mass-weighted self-shielding. The solid curves show the mass-weighted fractions, which deviate from the dashed volume-weighted ones during the EoR. In particular, because the first regions to be ionized are in larger overdensities, there is a bias arising from averaging over multi-phase ionization structure. This is most apparent for \HII (and hence electrons) around $z \sim 8$ and for \HeIII around $z \sim 3$.

\begin{figure}
    \centering
    \safeincludegraphics[width=\columnwidth]{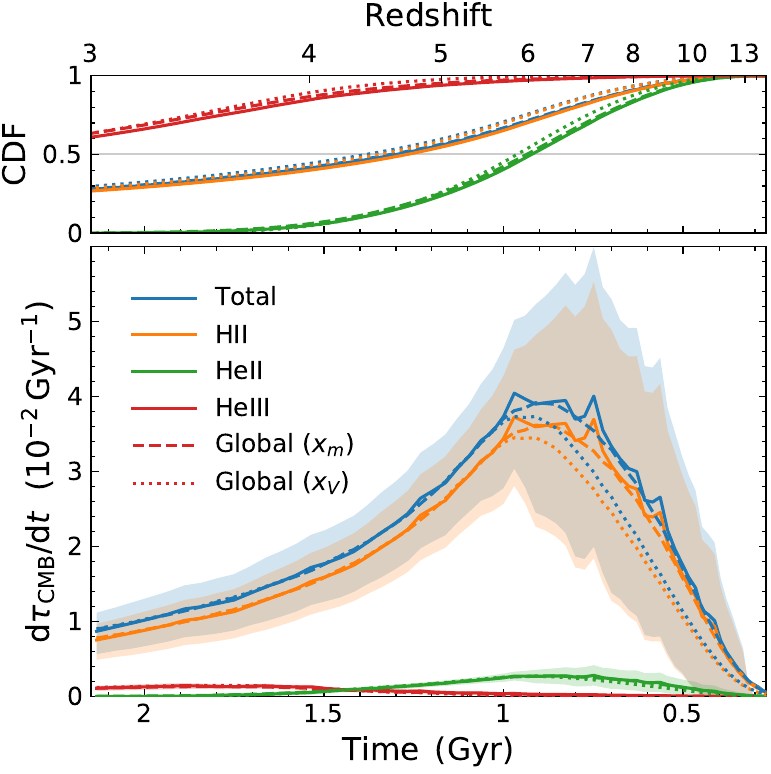}
    \caption{Normalized cumulative optical-depth histories (top panel) and differential optical depth $\text{d}\tau_\text{CMB}/\text{d}t$ (bottom panel) as a function of cosmic time (with the equivalent redshift shown above), based on global mean values and separated into contributions from \HII, \HeII, and \HeIII. The solid curves show the line-of-sight optical depth averaged over light-cone pixels, which follows the global mass-weighted average (dashed curves) but exceeds the global volume-weighted average (dotted curves) at all redshifts $z \gtrsim 6$. The shaded bands illustrate the pixel-to-pixel 1$\sigma$ confidence regions.}
    \label{fig:dtau_dt}
\end{figure}

In Fig.~\ref{fig:dtau_dt}, we propagate these ionization differences into the redshift evolution of the optical depth. The top panel shows cumulative optical-depth histories, where both the line-of-sight (LOS) curve computed by averaging the integrated optical depth along many sightlines and the mass-weighted global $\tau_\text{CMB}$ sit above the volume-weighted global $\tau_\text{CMB}$ during the EoR, and are in close agreement with each other. The bottom panel shows the differential contribution $\text{d}\tau_\text{CMB}/\text{d}t$, which peaks as soon as there are enough ionized bubbles and only drops when reionization is complete and the cosmic density decline takes over as $\rho \propto (1+z)^3$. Specifically, hydrogen (\HII) drives the main rise at $z \gtrsim 6$, singly-ionized helium (\HeII) contributes a smaller bump around the same time, and doubly-ionized helium (\HeIII) adds a modest tail at $z \lesssim 4.5$. This offset quantifies the patchiness bias discussed above where early ionized regions live in overdensities and the average integration along all sightlines is higher than would be expected from volume-weighting.

\subsection{Distribution of sightline optical depths}

\begin{figure}
    \centering
    \safeincludegraphics[width=\columnwidth]{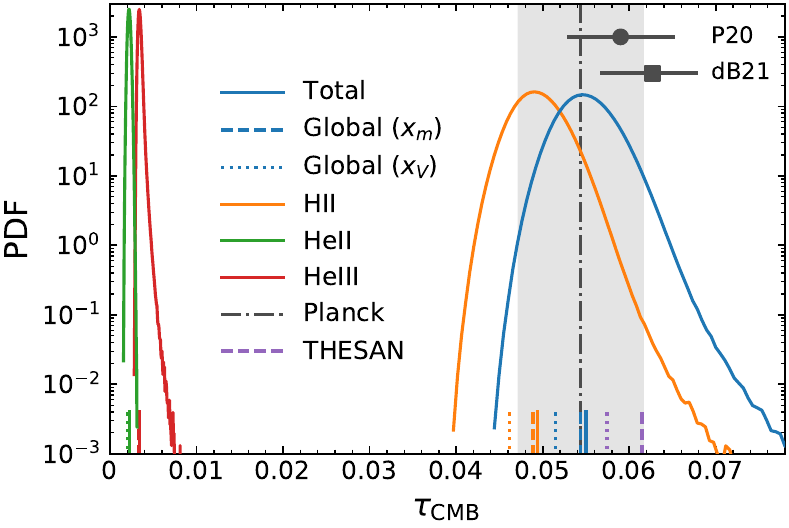}
    \caption{Histogram of $\tau_\text{LOS}$ values across all pixels of the light-cone map. For context, we include separate histograms for the total (all electrons), \HII, \HeII, and \HeIII contributions. The vertical dash-dotted line and gray $\pm1\sigma$ shaded band denote the Planck measurement \citep[$0.0544 \pm 0.0073$;][]{Planck2020} while markers show re-analysis by \citet[][P20]{pagano2020reionization} and \citet[][dB21]{de2021inference}. The vertical lines at the bottom show the light-cone (solid), mass-weighted (dashed), and volume-weighted (dotted) averages. Global values are also shown for \thesanone (purple) with higher values due to earlier reionization. The figure highlights both the dominance of \HII and the LOS enhancement relative to the standard global volume-weighted calculation.}
    \label{fig:tau_CMB_hist}
\end{figure}

In Fig.~\ref{fig:tau_CMB_hist}, we present the histogram of $\tau_\text{LOS}$ values across all pixels of the light-cone map, which reveals the one-point statistics of patchiness. The distribution is skewed with most sightlines having $\tau_\text{LOS}$ near the LOS average across pixels, but a long tail extends toward higher optical depths. The tail is populated by sightlines that pass through multiple overdense early ionized regions. Such extremes are rare enough to not significantly bias the mean, which is only larger than the median by $\langle \tau_\text{LOS} \rangle - \tilde{\tau}_\text{LOS} = 0.00013$ or approximately $0.23\%$. Therefore, the bias between the LOS average and the global volume-weighted mean is not driven by a small number of extreme rays. Instead, a high covering fraction of well-distributed bubbles across the sky ensures that the LOS integral averages over many ionized structures across redshift, producing a skewed but relatively narrow $\tau_\text{LOS}$ distribution compared to co-spatial integrations.

It is also clear that \HII dominates the total optical depth, while the \HeII and \HeIII distributions are too narrow to have a significant impact on the total variance, mostly due to the lower helium abundance and the timing of helium reionization. Effectively, \HeII traces \HII during the EoR, and despite the long time span over $z = 3$--$5$, cosmic expansion further dilutes $\tau_\text{CMB}$ differences arising from patchy helium reionization. Overall, the histogram builds intuition for later statistics, including the variance of the distribution and spatial correlations of fluctuations.

For comparison, the Planck 2018 constraints give $\tau_\text{CMB} \approx 0.0544 \pm 0.0073$ \citep[TT,TE,EE+lowE;][increasing to $0.056 \pm 0.0071$ with BAO]{Planck2020}, consistent with our LOS distribution. We also note that more recent re-analyses of the Planck HFI data obtain slightly higher values of $\tau = 0.059 \pm 0.006$ \citep{pagano2020reionization} and $\tau = 0.0627^{+0.0050}_{-0.0058}$ \citep{de2021inference}. These values are modestly higher than ours and would favor an earlier onset and/or a more extended high-redshift tail of reionization. Our global volume-weighted calculation ($\tau_{\text{CMB},V} = 0.0515$) lies below all of these values, but both the LOS light-cone average ($\langle \tau_\text{LOS} \rangle = 0.0550$) and global mass-weighted calculation ($\tau_{\text{CMB},m} = 0.0544$) are enhanced by density weighting and therefore agree more closely with the Planck range. This shows that patchiness can reduce the mismatch between a late volume-weighted simulation history and CMB constraints without invoking additional ionizing components. We note that related higher-resolution simulations such as \thesanone \citep{Kannan2022a} and \thesanhr \citep{Borrow2023} can result in an earlier gradual start to reionization from low-mass halos where star-formation is not sufficiently resolved in \lumina, so $\tau_\text{CMB}$ could increase further without significantly affecting late-time EoR observables. In fact, using the \thesanone reionization history at $z > 5.5$ and \lumina below gives $\tau_{\text{CMB},m}^\text{\thesan} = 0.0615$ and $\tau_{\text{CMB},V}^\text{\thesan} = 0.0574$, providing an illustrative high-end estimate for $\tau_\text{CMB}$ within this class of $\Lambda$CDM astrophysical models.

\begin{figure}
    \centering
    \safeincludegraphics[width=\columnwidth]{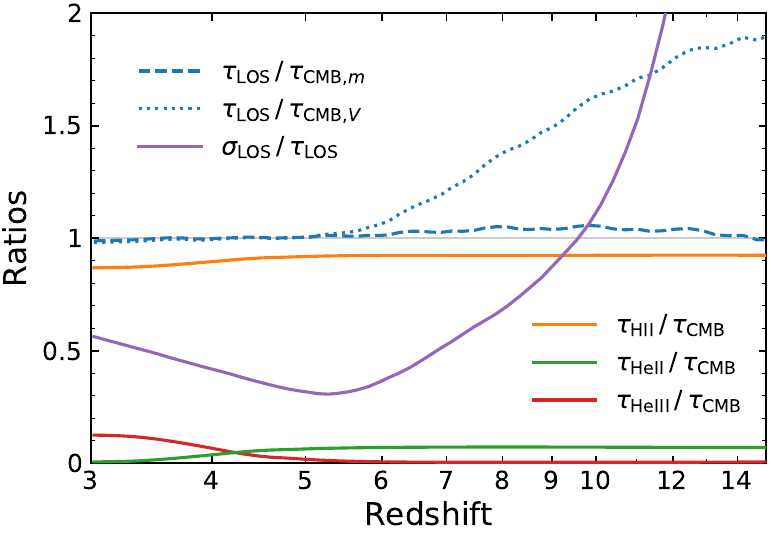}
    \caption{Redshift evolution of optical depth statistics, including the LOS average relative to the global mass- and volume-weighted averages (dashed and dotted curves), the standard deviation relative to the LOS average $\sigma_\text{LOS} / \langle \tau_\text{LOS} \rangle$, and fractional contributions of \HII, \HeII, and \HeIII. The LOS enhancement during the EoR is almost entirely accounted for by using mass-weighted ionized fractions, while the coefficient of variation shows that sightline fluctuations remain appreciable.}
    \label{fig:sigma_z}
\end{figure}

In Fig.~\ref{fig:sigma_z}, we show the redshift evolution of optical depth statistics, including the LOS average relative to the global mass- and volume-weighted averages (dashed and dotted curves), the standard deviation relative to the LOS average $\sigma_\text{LOS} / \langle \tau_\text{LOS} \rangle$, and fractional contributions of \HII, \HeII, and \HeIII. At low redshifts ($z \lesssim 6$), all sightlines converge to the global mean with $\langle \tau_\text{LOS} \rangle \approx \tau_{\text{CMB},V} \approx \tau_{\text{CMB},m}$. However, above $z \gtrsim 6$ this ratio starts to climb gradually, reaching a bias of nearly two by $z \approx 15$, which directly mirrors the $\langle x \rangle_m / \langle x \rangle_V$ boost rising at earlier times (see Fig.~\ref{fig:x_boost}). The coefficient of variation $\sigma/\tau$ reaches a minimum at $z \sim 5$--$6$ but overall remains modest ($\sim 0.5$) except at very high redshifts ($z \gtrsim 10$) when the optical-depth contribution is already diminishing in the simulation. Finally, the fractional contributions confirm that hydrogen dominates the optical depth across cosmic time with $\approx 10\%$ from helium. \HeII initially contributes at high redshifts but is overtaken by \HeIII by $z \sim 4$, when helium reionization is well underway. This breakdown is relevant for interpreting the physical origins of spatial fluctuations analyzed later between \HII, \HeII, and \HeIII arising from different epochs.

Overall, this supports the use of a single (mass-weighted) reionization history in many analyses, but precision measurements of CMB anisotropies can still probe early patchiness, especially when combined with other EoR timing- and topology-sensitive observables.

\begin{figure}
    \centering
    \safeincludegraphics[width=\columnwidth]{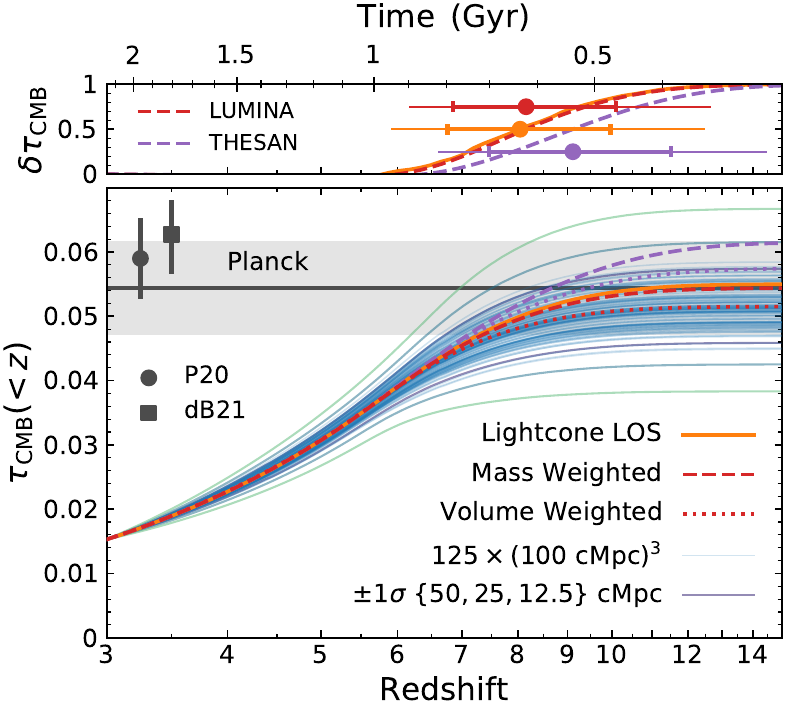}
    \caption{\textit{Top panel:} Relative excess of the light-cone signal to the volume-weighted average, $\delta\tau_\text{CMB} \equiv [\langle \tau_\text{LOS} \rangle(<z) - \tau_{\text{CMB},V}(<z)] / [\langle \tau_\text{LOS} \rangle -\tau_{\text{CMB},V}]$, highlighting that the LOS bias accumulates at $z_\text{LOS} = 8.04^{+1.93}_{-1.28}$, and likewise for the mass-weighted bias at $z_m = 8.15^{+1.96}_{-1.30}$. \textit{Bottom panel:} Cumulative optical depth $\tau_\text{CMB}(<z)$ for the light-cone LOS average (orange), global mass-weighted mean (red dashed), global volume-weighted mean (red dotted), 125 independent $100\,\text{cMpc}$ co-spatial subvolumes (thin curves), and the $\pm1\sigma$ variation between higher resolution co-spatial subvolumes exaggerating the possible ionized-density fluctuations. Planck constraints are also included and match the LOS and mass-weighted values. For comparison, \thesanone is included as purple curves.}
    \label{fig:cdf_z}
\end{figure}

\subsection{Cumulative optical depth and subvolume diversity}

In Fig.~\ref{fig:cdf_z}, the top panel quantifies when the LOS diverges most strongly from the global volume-weighted mean. We define the relative excess of the light cone as $\delta\tau_\text{CMB} \equiv [\langle \tau_\text{LOS} \rangle(<z) - \tau_{\text{CMB},V}(<z)] / [\langle \tau_\text{LOS} \rangle - \tau_{\text{CMB},V}]$. From this we find that the LOS bias accumulates at $z_\text{LOS} = 8.04^{+1.93}_{-1.28}$, and likewise for the mass-weighted bias at $z_m = 8.15^{+1.96}_{-1.30}$ (replacing $\langle \tau_\text{LOS} \rangle$ with $\tau_{\text{CMB},m}$), as well as for \thesanone with $z_m^\text{\thesan} = 9.11^{+2.41}_{-1.65}$, when the combined boost of a high cosmic density and the patchiness of ionized bubbles is maximal. This interval is close to the tanh-equivalent midpoint of reionization inferred by Planck ($z_\text{re} \approx 7.7 \pm 0.7$). The analysis therefore implies that patchy reionization can raise the effective $\tau_\text{CMB}$ by roughly $7\%$ while leaving the global volume-weighted history largely unchanged.

The bottom panel summarizes the cumulative optical depth $\tau_\text{CMB}(<z)$ as determined by light-cone sightline averages (orange), global mass-weighted (dashed) and volume-weighted (dotted) means, and co-spatial subvolumes (volume-weighted for consistency with coarsening) illustrating the cosmic variance of reionization histories at different resolutions. The comparison to CMB-inferred optical depths includes the homogeneous $z < 3$ contribution, $\tau_\text{CMB}(z < 3) = 0.0152$, which is not negligible for the mean (about $28\%$) but is common to all sightlines in our calculation. Residual fluctuations from the low-redshift cosmic web are not modeled here and would enter as a further sub-percent contribution to the sightline scatter rather than changing the mean optical depth. Each of the 125 thin transparent curves corresponds to a different $100\,\text{cMpc}$ region. Their spread with increasing redshift shows how widely the local reionization midpoint and duration can vary. Some regions reionize early and produce a high cumulative $\tau_\text{CMB}$, while others lag and produce a low value. On average, the subvolume curves recover the global history, but the diversity is substantial, especially considering that previous large-volume radiation-hydrodynamical reionization simulations are each of comparable volume albeit calibrated to still match the observed reionization history \citep{Kannan2022a, ocvirk2020cosmic, croc120cMpc}. The LOS and mass-weighted curves sit above most volume-weighted subvolumes, reflecting the bias when accumulating optical depth from early ionized overdensities. This behavior is consistent with related \thesan analyses showing that reionization timing depends strongly on large-scale context \citep{Zhao2026}.

\begin{figure}
    \centering
    \safeincludegraphics[width=\columnwidth]{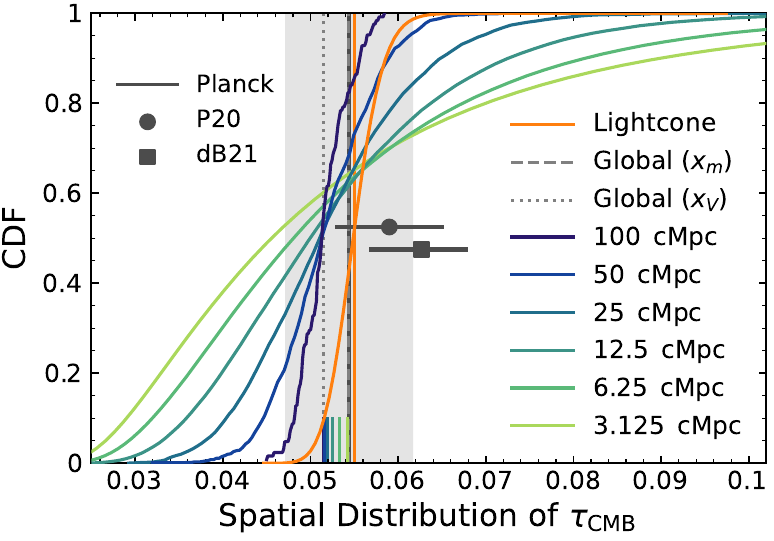}
    \caption{Cumulative distribution functions of $\tau_\text{CMB}$ calculated from local reionization histories for subvolumes of different sizes ranging from $\{100, 50, 25, 12.5, 6.25, 3.125\}\,\text{cMpc}$. The sampled variance increases with resolution and in comparison the light-cone LOS distribution (orange curve) has a significantly narrower range. Vertical lines and markers indicate the global volume-weighted (gray dotted) and mass-weighted (gray dashed) averages, increasing co-spatial averages, and various Planck constraints.}
    \label{fig:hist_local}
\end{figure}

In Fig.~\ref{fig:hist_local}, we extend the previous co-spatial analysis to even smaller scales. As the subvolume size decreases, the distribution of $\tau_\text{CMB}$ shifts toward higher values and broadens considerably. At $100\,\text{cMpc}$, comparable to previous simulation volumes, the mean remains centered on the global volume-weighted average (dotted gray line) but the variance is already at the level of the LOS distribution from the high-resolution light cone (orange curve). Continuing to decrease by factors of two until $3.125\,\text{cMpc}$ (for $160^3$ subvolumes), which is below the scale where most ionized bubbles are resolved, the distribution has developed an extended high-$\tau$ tail and the mean has reached the global mass-weighted average (dashed gray line). This progression shows that the clumping bias becomes resolved as one probes smaller regions, the high-density, early-reionized patches are represented without being washed out into their surrounding lower-density regions, and the mean converges toward the mass-weighted average.

\begin{figure}
    \centering
    \safeincludegraphics[width=\columnwidth]{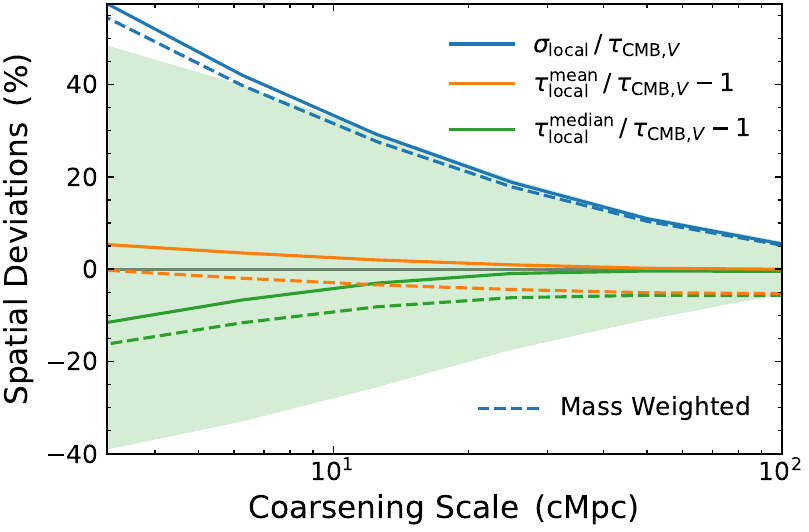}
    \caption{Spatial standard deviation of $\sigma_\text{local}$ (blue) relative to the global volume-weighted ($\tau_{\text{CMB},V}$; solid) and mass-weighted ($\tau_{\text{CMB},m}$; dashed) averages as a function of the coarsening scale ranging from $3.125$--$100$\,cMpc subvolumes. The mean (orange) and median (green) curves show that averaging over too large a region ($\sim 30$\,cMpc) results in similar values for the mean (orange) and median (green), both of which are biased low. Resolving smaller scales produces a divergence in these quantities with the mean increasing and the median decreasing. The bias is finally removed once bubbles are sufficiently resolved ($\sim 3$\,cMpc). The shaded region indicates the $\pm1\sigma$ spread.}
    \label{fig:sigma_local}
\end{figure}

As a final summary, in Fig.~\ref{fig:sigma_local} we show the spatial standard deviation $\sigma_\text{local}$ (blue) and relative deviation of the mean (orange) and median (green) of the distribution across subvolumes from the global volume-weighted (solid curves) and mass-weighted (dashed curves) averages. As the coarsening scale decreases from $100$\,cMpc to $3.125$\,cMpc, the coefficient of variation ($\sigma/\tau$) increases significantly, reaching nearly 50\% and likely continues to rise as ever higher densities are accessible within the filtering scale. The mean and median are in agreement when the coarsening scale is $\gtrsim 30$\,cMpc, below this the mean increases to the global mass-weighted value while the median drops significantly. This divergence reflects the skewness seen in the previous histograms in Fig.~\ref{fig:hist_local} in which a decreasing fraction of high-$\tau$ subvolumes carries the mean, whereas most subvolumes represent lower-density, later-reionized regions that remain below the global mean. The volume-weighted bias is finally removed once bubbles are sufficiently resolved ($\sim 3$\,cMpc). The $\pm1\sigma$ variation (shown by the green shaded region) also grows steadily with decreasing scale, meaning that more localized regions exhibit more scatter in their reionization histories. These results suggest that predicting $\tau_\text{CMB}$ variations from a limited simulation volume or interpreting observed CMB anisotropies requires accounting for spatial and light-cone sightline variance giving rise to a non-Gaussian distribution of optical depths.

\begin{figure}
    \centering
    \safeincludegraphics[width=\columnwidth]{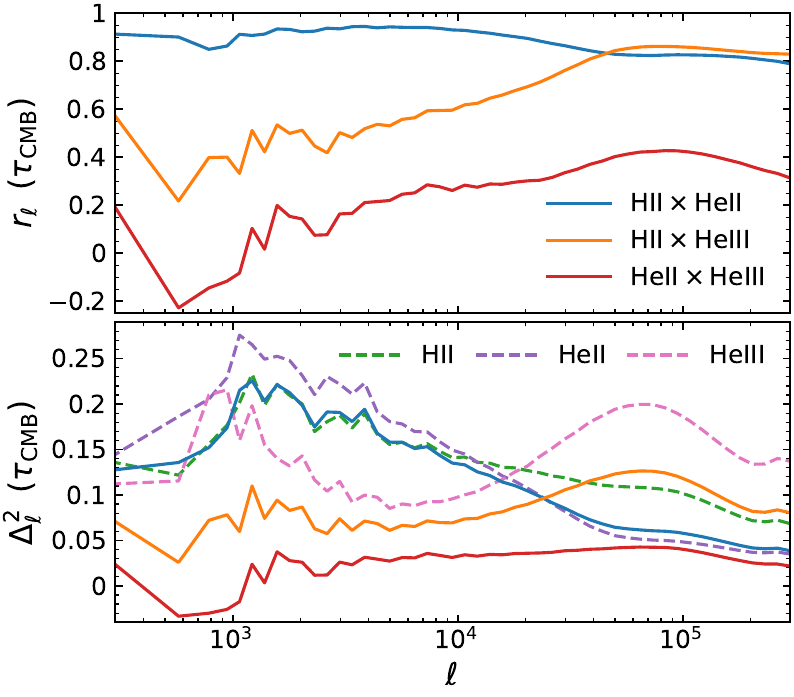}
    \caption{The angular cross-correlation coefficient $r_\ell \equiv C_\ell^{xy} / \sqrt{C_\ell^{xx} C_\ell^{yy}}$ (top panel) and angular power spectra (bottom panel) of the light-cone $\tau_\text{LOS}$ map, separated into \HII, \HeII, and \HeIII contributions. To emphasize scale features, we show the dimensionless power $\Delta_\ell^2 \equiv \ell(\ell+1)\,C_\ell/(2\pi)$, with auto-spectra (dashed curves) and cross-spectra (solid curves). All maps have been mean-subtracted and normalized to unit variance, so this figure compares spectral shapes rather than absolute amplitudes. Multipoles near the field-of-view scale, $\ell_\text{cone} \approx 100$, are the most sensitive to retiling and finite-area effects.}
    \label{fig:spectrum}
\end{figure}

\subsection{Angular and real-space statistics}

Finally, we characterize the angular structure of $\tau_\text{LOS}$ fluctuations and the extent to which hydrogen and helium contributions trace one another. In the top panel of Fig.~\ref{fig:spectrum}, we show the cross-correlation coefficient $r_\ell \equiv C_\ell^{xy} / \sqrt{C_\ell^{xx} C_\ell^{yy}}$. \HII and \HeII are very well correlated ($r_\ell > 0.8$) across all $\ell$, with the cross-spectra selecting the shape of the lower of the two. \HII--\HeIII shows strong correlations on small scales ($\ell \gtrsim 4 \times 10^4$) but declines on larger scales as the light cone traverses different regions of the universe during hydrogen and helium reionization, and therefore a transfer of structure. Finally, the \HeII--\HeIII cross-spectrum is nearly featureless, although the cross-correlation reveals modest correlation on small scales and an anti-correlation on the largest trustworthy scales ($\ell_\text{cone} \approx 100$). This arises because \HeII is converted into \HeIII at later times, so their spatial distributions only partially overlap during the epoch of helium reionization comprising the entirety of the \HeIII signal ($z \lesssim 5$; see Fig.~\ref{fig:dtau_dt}).

The bottom panel presents the radially-averaged power spectra of fluctuations to reveal the characteristic angular scales of optical depth fluctuations. To more easily compare the angular behavior of fluctuation shapes rather than amplitudes, all maps are mean-subtracted and normalized to have unit variance, and we show the dimensionless power $\Delta_\ell^2 \equiv \ell(\ell+1)\,C_\ell/(2\pi)$. The $\tau_\text{LOS}$ auto-spectrum (dashed curves) peaks at multipoles of $\ell \sim 1$--$5 \times 10^3$ for \HII and \HeII, corresponding to angular scales of roughly $4.3$--$22$\,arcminutes or $8.3$--$50$\,cMpc at $z = 5$. At higher multipoles the power drops, reflecting the finite size of ionized bubbles and the smoothing effect of line-of-sight integration. The cross-spectra for \HII--\HeII, \HII--\HeIII, and \HeII--\HeIII bring out more strongly the differences in scale-dependent physics. In the unnormalized maps, the \HII spectrum dominates at all scales, with \HeII appearing similar but with a steeper slope towards smaller scales because of the higher ionization threshold compared to \HII and the eventual conversion to \HeIII, making helium structures less coherent on small scales. On the other hand, \HeIII exhibits strong power on multipoles of $\ell \sim 0.6$--$2 \times 10^3$ and $\sim 0.2$--$2 \times 10^5$, corresponding to angular scales of roughly $11$--$36$ and $0.1$--$1$\,arcminutes or $25$--$83$ and $0.25$--$2.5$\,cMpc at $z = 5$, respectively. The bimodal prominence of power at both low- and high-$\ell$ illustrates a difference in the topology of hydrogen and helium reionization, with rare AGN driving the largest \HeIII bubbles, and fainter AGN also creating a network of relatively small structures throughout the box. The LOS integration appears to be enough to suppress shot noise.

\begin{figure}
    \centering
    \safeincludegraphics[width=\columnwidth]{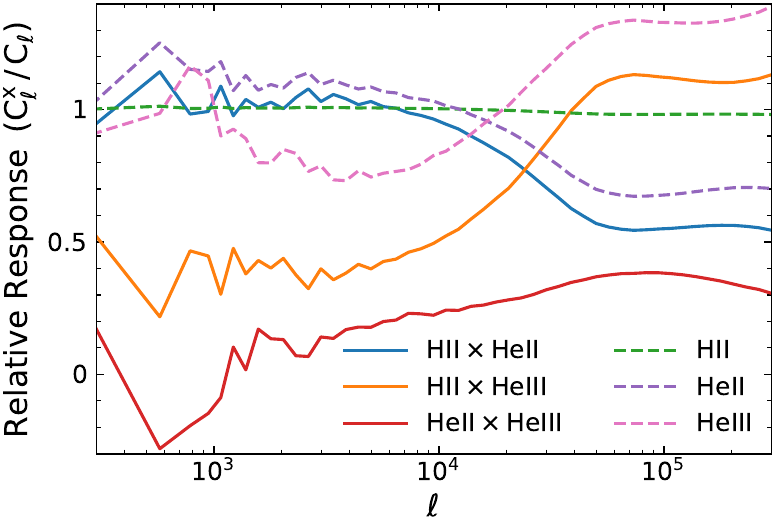}
    \caption{Scale-dependent response of each auto- and cross-component relative to the total $\tau_\text{LOS}$ spectrum. For auto terms the plotted quantity is $\sqrt{C_\ell^x/C_\ell}$ and for cross terms it is $C_\ell^{xy}/C_\ell$ to allow negative signals. Unity indicates perfect proportionality to the total spectrum shape.}
    \label{fig:spectrum_bias}
\end{figure}

In Fig.~\ref{fig:spectrum_bias}, we compress the spectral information into a scale-dependent bias or relative response function. Specifically, for auto terms we show $\sqrt{C_\ell^x/C_\ell}$ and for cross terms $C_\ell^{xy}/C_\ell$ to allow negative signals. The clearest trend is that \HII (dashed green curve) lies close to unity across all scales, confirming that hydrogen sets the overall shape. \HeII (dashed purple) and \HeIII (dashed pink) have mild opposing biases, each being flat and featureless at $\ell \gtrsim 5 \times 10^4$ with values that are $\sim 30\%$ lower (higher) for \HeII (\HeIII). However, at larger scales, there is a crossover ($\ell \sim 1$--$3 \times 10^4$) with the power being comparatively overrepresented (suppressed) for \HeII (\HeIII). We interpret this as an overall reshaping due to the \HeII-to-\HeIII transformation such that \HeII fluctuations are a smoothed version of the \HII field. The recovery of \HeIII at both large and small scales may also reflect the time lag and different bubble structure between hydrogen and helium reionization. Cross terms mirror these trends as \HII--\HeII (blue) is an unbiased tracer on large scales but drops to half the power over $\ell \approx 0.7$--$5 \times 10^4$. Meanwhile, \HeII--\HeIII (red) remains low across all scales, reaching a maximum ratio below $0.4$ at $\ell \sim 10^5$ and a minimum (even an anti-bias) at $\ell \approx 600$. Finally, \HII--\HeIII (orange) shows a strongly diverging bias of high values on small scales and low ones on large scales. Although not directly separable in observations, the physically informative pairings show that helium ionization states (especially \HeIII) introduce scale-dependent deviations from the hydrogen morphology.

\begin{figure}
    \centering
    \safeincludegraphics[width=\columnwidth]{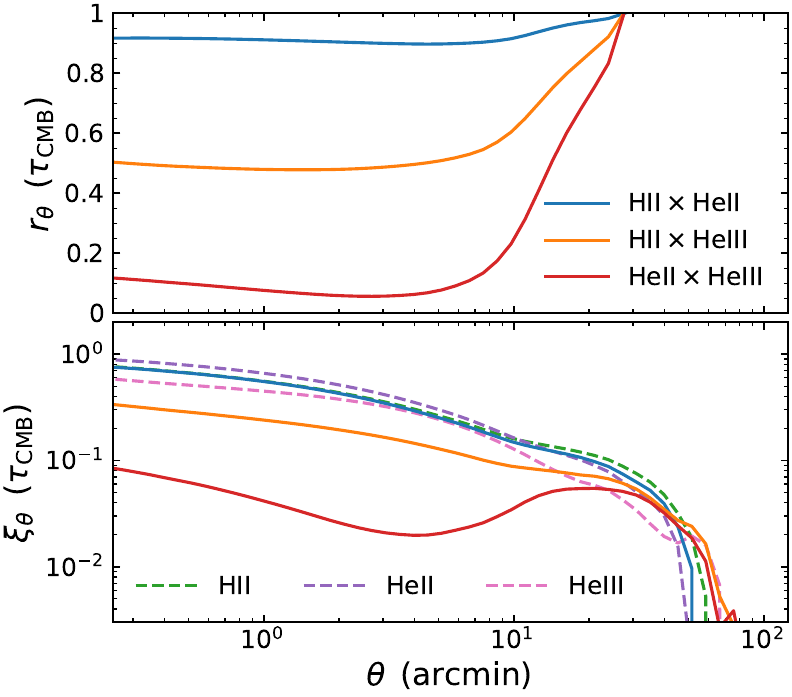}
    \caption{\textit{Top panel:} The cross-correlation coefficient $r_\theta \equiv \xi_\theta^{xy} / \sqrt{\xi_\theta^{xx} \xi_\theta^{yy}}$, highlighting structural differences between pairings of \HII, \HeII, and \HeIII. \textit{Bottom panel:} Two-point correlation functions $\xi(\theta)$ of $\tau_\text{LOS}$ light-cone maps, also separated into components of \HII, \HeII, and \HeIII. Dashed curves denote auto-correlations and solid lines are cross-correlations. Like the angular power spectra, the maps are mean-subtracted and normalized to unit variance to emphasize the shape as a function of angular separation $\theta$.}
    \label{fig:corr}
\end{figure}

For a complementary picture directly in real space, the top panel of Fig.~\ref{fig:corr} shows the angular cross-correlation coefficient, $r_\theta \equiv \xi_\theta^{xy} / \sqrt{\xi_\theta^{xx} \xi_\theta^{yy}}$. Specifically, \HII--\HeII (blue) exhibits a strong correlation ($r_\theta > 0.9$) on all scales and the $\xi$ shape rests between the individual ones, indicating near-identical morphologies. The correlation function for \HII--\HeIII is a factor of two weaker than the one for \HII--\HeII, with $r_\theta \approx 0.5$ until $\sim 10'$ separations, after which the largest scales are once again correlated. This suggests similar structure for $\tau_\text{LOS}$ at separations large enough to sample early-reionized regions and growing bubbles. Finally, the \HeII--\HeIII cross-correlation function is suppressed by an order of magnitude, and indeed $r_\theta \lesssim 0.1$ until the characteristic $\sim 10'$ coherence scale. Still, the re-emergence of coherence at scales larger than this is comparable to other cross-correlations, suggesting the percolation process of helium bubbles (and the associated \HeII to \HeIII conversion) is relatively generic.

The bottom panel shows the two-point correlation functions $\xi(\theta)$ of $\tau_\text{LOS}$ calculated from the light-cone maps as a function of angular separation $\theta$. We show both auto-correlations (dashed curves) separated into components of \HII, \HeII, and \HeIII, as well as cross-correlations (solid curves) for the same pairings. As we are emphasizing the shape and scale-dependence, all maps are mean-subtracted and normalized to unit variance. The \HII correlation function (green) sets the dominant coherence scale, which begins at the smallest trustworthy scales (one pixel is approximately $2.5''$) and falls smoothly with separation, dropping by a factor of five between $\sim 10''$ and $\sim 10'$, corresponding to $\sim 0.4$--$25$\,cMpc at $z = 5$. Beyond this, some degree of correlation is retained until a sharp cutoff at $\sim 1^\circ$, due to the limited field of view ($3.6^\circ$). The \HeII auto-correlation (purple) is slightly stronger (weaker) on smaller (larger) scales, which is likely a difference in the $z \lesssim 6$ behavior. \HeIII (pink) shows a suppression at all scales, especially above $\sim 10'$ due to the different topology of helium reionization.

\begin{figure}
    \centering
    \safeincludegraphics[width=\columnwidth]{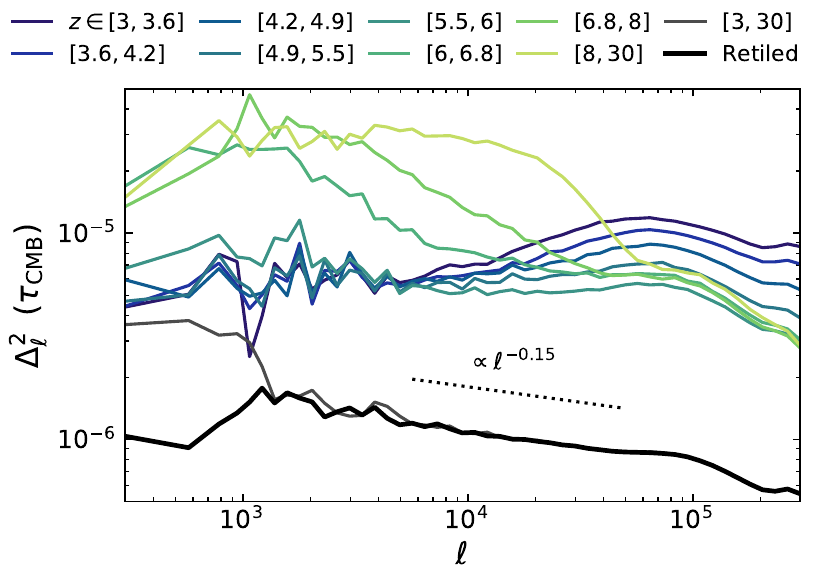}
    \caption{Dimensionless angular power spectra $\Delta_\ell^2 \equiv \ell(\ell+1)\,C_\ell/(2\pi)$ of $\tau_\text{LOS}$ fluctuations separated into equal-$\tau$ redshift ranges as indicated (following Fig.~\ref{fig:lightcone_image_ranges}). The vertical scale reflects the absolute fluctuation power. The total $z \in [3,30]$ spectrum is shown by the gray curve, while the black curve uses our fiducial retiled light cone; these total curves are the absolute simulated $C_\ell^{\tau\tau}$ signals, shown in dimensionless form, relevant for comparison to patchy-$\tau$ reconstruction forecasts. The dotted line illustrates a simple power-law fit $\propto \ell^{-0.15}$. Slices at $z \gtrsim 6$ exhibit larger low-$\ell$ power because of the rich structure of clustered ionized regions, while slices at $z \lesssim 5$ produce a flatter $\Delta_\ell^2$, reflecting the smoother post-reionization Universe. As in Fig.~\ref{fig:spectrum}, the lowest multipoles are most sensitive to the finite field of view and retiling choice.}
    \label{fig:spectrum_z}
\end{figure}

The correlation functions indicate that $\tau_\text{LOS}$ is statistically stable at the resolutions studied. Because the scattering integral scales linearly with electron density, small-scale clumping (below $\sim 100$\,ckpc) can only contribute modestly to the optical depth, unlike recombination-sensitive quantities that scale as density squared. Recombination-sensitive quantities can be sensitive to: (\textit{i}) resolution, with additional clumping on sub-grid scales, or (\textit{ii}) assumptions in the physical modeling. The correlation functions thus suggest that our results are relatively insensitive to further unresolved small-scale fluctuations.

To complement the species-based analysis, we also investigate how the EoR affects the structure of the optical-depth fluctuations. In Fig.~\ref{fig:spectrum_z}, we use the same redshift ranges as in Fig.~\ref{fig:lightcone_image_ranges}, which divide $z \in [3,30]$ into eight bins with equal contributions to $\tau_\text{CMB}$, and calculate the angular power spectra and correlation functions for each bin. For each slice, the map is re-scaled so that its mean optical depth matches that of the full light cone, i.e.\ we preserve the absolute power rather than normalizing to unit variance, and then the mean is subtracted to isolate fluctuations. Thus, unlike Fig.~\ref{fig:spectrum}, the total spectrum in Fig.~\ref{fig:spectrum_z} is the simulated $C_\ell^{\tau\tau}$ signal (displayed as $\Delta_\ell^2$) most directly comparable to patchy-$\tau$ reconstruction noise forecasts.

\begin{figure}
    \centering
    \safeincludegraphics[width=\columnwidth]{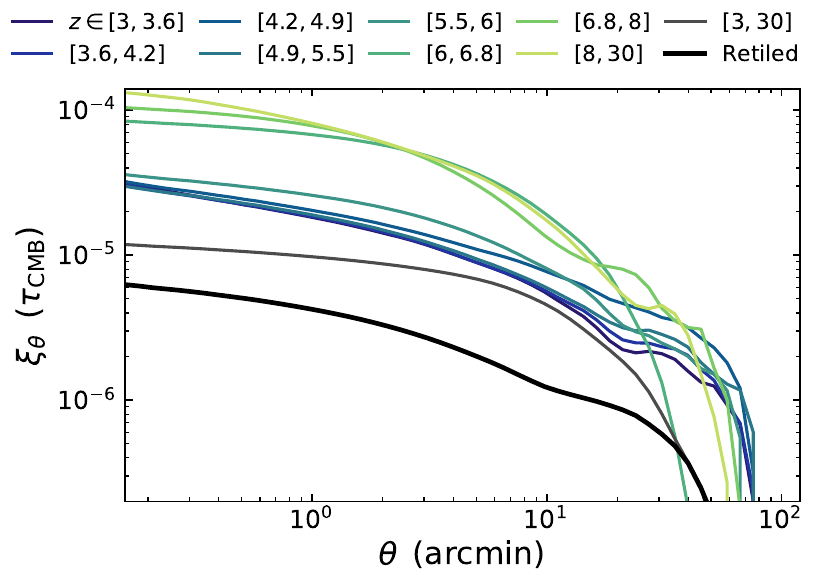}
    \caption{Redshift-sliced real-space correlation functions of $\tau_\text{LOS}$. The two-point correlation function $\xi_\theta$ is plotted as a function of angular separation for the same equal-$\tau$ redshift bins as in Figs.~\ref{fig:spectrum_z} and \ref{fig:lightcone_image_ranges}. The curves reflect the absolute correlation amplitude of fluctuations. Early slices ($z \gtrsim 6$) display strong coherence at separations $\theta \lesssim10'$, while later slices flatten rapidly, indicating a nearly uniform optical depth once hydrogen reionization has completed. The total $z \in [3,30]$ correlation (gray curve) shows that the redshift integration reduces spatial coherence on small scales, while the light-cone retiling procedure (black curve) further suppresses large-scale coherence, with the fiducial case lying between these limits.}
    \label{fig:corr_z}
\end{figure}

These slices reveal a clear evolution in the fluctuation amplitude. At $z \gtrsim 6$ the Universe is still highly patchy, and the corresponding curves dominate the small-scale power ($\ell \gtrsim 10^4$), while at lower redshifts, the IGM is more homogeneous, resulting in flatter $\Delta_\ell^2$. A simple power-law fit to the total (gray) curve gives $\Delta_\ell^2 \propto \ell^{-0.15}$, emphasized by the dotted line. The retiled light cone (black) has nearly the same shape but shows a suppression of power at $\ell \lesssim 10^3$ due to the variance minimizing strategy to avoid repeated structures, although retiling does not affect higher multipoles. These redshift-channeled spectra show that during reionization strong low-$\ell$ correlations arise from the characteristic ionized bubble size and separation scales, while cosmic structure in the post-reionization Universe imprints more high-$\ell$ power.

The real-space analog is shown in Fig.~\ref{fig:corr_z}. Here we plot the two-point correlation function $\xi_\theta$ for the same redshift slices, again preserving the full mean and then subtracting it to focus on fluctuations. At separations $\theta \lesssim 10'$ the correlation is strongest for the earliest slices, reflecting the small but dense ionized patches present at high redshift. The amplitude declines with decreasing redshift, flattening significantly once reionization is complete. All curves also fall sharply at $\theta \sim 1^\circ$, at which point the fluctuations are no longer coherent. The retiled correlation function (black) is again depressed relative to the original (gray), confirming that the tiling procedure primarily suppresses large-scale variance without strongly altering the structure. These results reinforce the conclusion that patchy high-redshift structure drives observed fluctuations, while the late-time IGM contributes relatively little residual variation.

\section{Discussion}
\label{sec:discussion}

Our results emphasize that $\tau_\text{CMB}$ is not only an integral constraint on the reionization history but also a diagnostic of how ionization correlates with large-scale structure. In \lumina, the $\approx 7\%$ enhancement of the light-cone average $\langle \tau_\text{LOS} \rangle$ over a calculation based on the global volume-weighted ionization history is reproduced by the global mass-weighted history. The dominant correction arises from the density--ionization correlation in which ionized regions form first in overdensities with disproportionately large electron columns. Since CMB analyses already constrain the mean optical depth, this only affects the corresponding mapping to volume-weighted simulation proxies. Interpreting $\tau_\text{CMB}$ as a volume-averaged ionized fraction (or volume filling factor) can lead to inconsistencies or systematic biases when combined with observations of individual sightlines, e.g.\ the Ly$\alpha$ forest. Previous analyses explicitly incorporate density weighting \citep{Trac2018} but this distinction is not always made explicit in observationally motivated parameterizations and in some simulation post-processing pipelines. We provide an analytic fit, calibrated on \lumina, for converting between $\langle x \rangle_m$ and $\langle x \rangle_V$ in empirical applications with similar ionization histories (see Fig.~\ref{fig:x_boost}). However, our coarse-graining experiments show that the correction itself depends on resolving bubble-scale structure, because smoothing on $\gtrsim 3$\,cMpc scales suppresses the correlation and biases $\tau_\text{CMB}$ low.

The mean optical-depth correction is less sensitive to the detailed appearance of any particular map than the higher-order statistics are. Retiling changes which structures are juxtaposed on the sky and therefore affects the variance, tails, and lowest angular multipoles, but it does not change the mean LOS optical depth. In this sense, the $\approx 7\%$ offset between the volume-weighted and light-cone means is a density-weighting effect, while the exact value of $\sigma_\text{LOS}$ and the low-$\ell$ power should be interpreted with the finite field of view and tiling strategy in mind (Appendix~\ref{ap:retiling}). The resolution tests further show that the mean converges once the relevant bubble-scale density--ionization structure is resolved (Appendix~\ref{ap:resolution}).

The predicted LOS dispersion, $\sigma_\text{LOS} / \langle \tau_\text{LOS} \rangle \gtrsim 5.0\%$ ($6.5\%$ without retiling), shows that higher-order information is contained in the full spatially-varying $\tau_\text{LOS}$ field beyond the average. Conceptually, this variance reflects the stochastic presence of early-ionized overdensities and large-scale structure. While the primary CMB inference of $\tau_\text{LOS}$ is an all-sky measurement, patchiness induces additional anisotropy signatures that encode the underlying morphology of screening, scattering, and kSZ effects \citep{Dvorkin2009Bmode, Zahn2012, Battaglia2013, SmithFerraro2017, Chen2023_PatchykSZ, Kramer2025}. In particular, \citet{Dvorkin2009Bmode} showed that the patchy-screening contribution to reionization-generated B-mode polarization dominates over the scattering contribution at $\ell \gtrsim 300$, making the optical-depth fluctuations measured here relevant for polarization anisotropies as well as for reconstruction of anisotropic optical depth \citep{dvorkin2009reconstructing}. High-resolution spatial maps also carry the imprint of the bubble size distribution and other topological information \citep{Neyer2024, Neyer2026}. Indeed, the angular power spectra and correlation functions, including the nontrivial behavior of helium cross-correlations (Figs.~\ref{fig:spectrum}--\ref{fig:corr}), provide additional insights connecting the simulated reionization morphology to EoR observables.

These connections matter for cosmology because $\tau_\text{CMB}$ suppresses the primary CMB temperature and polarization spectra and is therefore degenerate with the primordial fluctuation amplitude. An attractive possibility is to use observables that trace the patchy EoR more directly to infer the same mean optical depth. For example, the kSZ signal measures the Doppler imprint of moving free electrons, and forecasts combining its power spectrum with higher-order information suggest that CMB-S4-like data could constrain both the mean optical depth and the duration of reionization \citep{Alvarez2021}. Similarly, 21\,cm power spectra and global-signal measurements probe the neutral gas that complements the ionized regions, and have been proposed as ways to predict $\tau_\text{CMB}$ independently of large-angle CMB polarization \citep{Liu2016}. Our results support the idea that reionization morphology contains information about the mean optical depth, but this is not simply a conversion from fluctuation amplitude to $\langle \tau_\text{LOS} \rangle$. The skewed $\tau_\text{LOS}$ distribution, the divergence between mean and median on small scales, the mass- versus volume-weighted boost, and the scale-dependent H/He contributions all indicate that overly restrictive patchiness templates could bias such inferences. Robust use of kSZ, 21\,cm, or reconstructed patchy-$\tau$ measurements should therefore forward-model not only the average $C_\ell^{\tau\tau}$ amplitude but also the covariance, non-Gaussianity, and relation between fluctuation observables and the mean optical depth.

Large-angle CMB polarization also contains some information about when reionization occurred beyond the total optical depth. Many analyses compress the ionization history into a tanh-like transition, as implemented for example in CAMB, with a midpoint chosen to reproduce a desired $\tau_\text{CMB}$ \citep{Lewis2008}. This is useful, but histories with the same integrated optical depth can produce slightly different low-$\ell$ E-mode ``reionization bumps'' if their ionization is distributed differently in redshift. Principal-component approaches describe this additional information in a model-independent way \citep{Hu2003, Mortonson2008}. The global electron-weighted histories from \lumina could therefore be passed directly to a Boltzmann calculation to predict the reionization bump, rather than reducing the simulation to a single integrated $\tau_\text{CMB}$. A direct likelihood analysis of this kind is beyond the scope of the present work and would likely be limited by cosmic variance and foreground systematics, but it provides another route for comparing physically motivated reionization histories to CMB data.

The Planck reionization analysis provides additional context for these comparisons. Although the baseline Planck optical-depth constraint uses a simple tanh reionization model, tests with more flexible PCA and FlexKnot histories give broadly consistent values of $\tau_\text{CMB}$, suggesting that current large-scale polarization constraints are fairly robust to the assumed parametrization \citep{Planck2020}. At the same time, these flexible reconstructions favor a late and relatively rapid transition, with little evidence for a large high-redshift contribution to the optical depth. Thus, modest corrections from patchiness and mass weighting are important for precision comparisons, while much larger optical depths would require unusually extended or early ionization histories. Planck also emphasizes that patchy kSZ limits are not straightforward to translate into a reionization duration without assumptions about the ionization topology, which further motivates simulation-based modeling when connecting patchy observables to a mean optical depth.

Finally, the magnitude and redshift range of the light-cone boost clarify how patchiness interfaces with recent discussions of potentially higher values for $\tau_\text{CMB}$. The boost accumulates near $z \sim 8$ (Fig.~\ref{fig:cdf_z}), when the density weighting in Eq.~(\ref{eq:opticalDepthCMB}) is large and the IGM is maximally patchy. Updating from volume to mass weighting and applying the helium reionization history from \lumina revises the previously reported \thesanone value of $\tau_{\text{CMB},V}^\text{\thesan} = 0.0574$ \citep{Kannan2022a} up to $\tau_{\text{CMB},m}^\text{\thesan} = 0.0615$, in line with re-analysis values of $\tau = 0.059$ \citep{pagano2020reionization} and $\tau = 0.0627$ \citep{de2021inference}, while \lumina matches the Planck 2018 value \citep[$\tau = 0.0544$;][]{Planck2020}. This perspective reduces the mismatch between volume-weighted late reionization histories and current CMB constraints, but the effect is much smaller than the change required by the high-$\tau_\text{CMB}$ cosmology scenario. Raising the optical depth from $\sim0.06$ to $\sim0.09$ is a $\sim50\%$ increase, far larger than the $\approx 7\%$ correction from volume weighting. Such a value, explored by \citet{Sailer2026} as a way to ease DESI BAO--CMB tensions in $\Omega_m$ and related inferences for dark energy and neutrino mass, would require substantially more ionization at early times than is produced by this class of astrophysical reionization models, while still remaining consistent with late-time EoR constraints. Similarly, \citet{Dai2026} found that higher-$\tau$ reionization histories can shift recent BAO+CMB neutrino-mass fits toward positive $\sum m_\nu$, with the effect driven mainly by the total optical depth rather than the detailed history. If future analyses continue to explore higher-$\tau$ solutions \citep{Giare2024, Elbers2025, Sailer2026, Kageura2026}, our results motivate treating the patchiness correction and the choice of mass versus volume weighting as explicit, physically motivated ingredients of comparisons between simulations, semi-analytic models, and CMB-inferred optical depths.

\section{Conclusions}
\label{sec:conclusions}

We have used the novel \lumina radiation--hydrodynamical simulation \citep{Zier2026}, which adopts the IllustrisTNG galaxy-formation model at high resolution ($2 \times 6000^3$ elements) in a large volume ($L_\text{box} = 500$\,cMpc), to study the CMB Thomson optical depth as both a global constraint and a spatial field, leveraging on-the-fly light cones and high-cadence Cartesian outputs that track \HII, \HeII, and \HeIII through hydrogen and helium reionization down to $z = 3$. Our main conclusions are:
\begin{enumerate}[leftmargin=*]
  \item Patchiness boosts the mean optical depth relative to volume-weighted histories. Specifically, we calculate volume-weighted, mass-weighted, and light-cone-integrated values of $\tau_{\text{CMB},V} = 0.0515$, $\tau_{\text{CMB},m} = 0.0544$, and $\langle \tau_\text{LOS} \rangle = 0.0550$, respectively. The light-cone value is $\approx 7\%$ above the volume-weighted proxy but only $\approx 1\%$ above the mass-weighted value, which is the closer analog of the electron column constrained by CMB analyses. Mass-weighted ionization histories therefore capture nearly the full correction, as the excess optical depth is driven by an early overdensity--ionization correlation at $z \gtrsim 6$, with a peak patchy signal near $z \sim 8$.
  \item $\tau_\text{LOS}$ exhibits substantial, non-Gaussian sightline dispersion; even with a variance-minimizing retiling strategy, the standard deviation is $\sigma_\text{LOS} = 0.00277$ (or $\approx 5.0\%$ of the mean) compared to $\sigma_\text{LOS} = 0.00359$ (or $\approx 6.5\%$ of the mean) without retiling, with a skewed distribution whose tails arise from rare, early-ionized overdensities.
  \item Optical depth predictions are sensitive to the averaging scale. Coarse-graining on $\gtrsim 3$\,cMpc scales suppresses the density--ionization correlation and biases $\tau_\text{CMB}$ low, while resolving bubble-scale structure ($\lesssim 3$\,cMpc) recovers the mass-weighted mean.
  \item Angular statistics show nontrivial hydrogen--helium relationships and redshift evolution of pre- and post-EoR contributions. Angular power spectra and correlation functions of $\tau_\text{LOS}$ decomposed into \HII, \HeII, and \HeIII contributions exhibit strong \HII--\HeII coherence but scale-dependent departures compared to \HeIII, reflecting the distinct topology of AGN-driven helium reionization.
\end{enumerate}

This paper serves as an early showcase of the \lumina light-cone framework. Although we have only discussed derived $\tau_\text{CMB}$ constraints, we anticipate that future studies will extend these analyses to additional reionization observables \citep[e.g.\ tSZ/kSZ, 21\,cm--LAE cross-correlations, and line-intensity mapping;][]{Bernal2022, Kannan2022b, Qin2022, Iliev2025, Almualla2025, Chen2026} and will forward-model CMB temperature and polarization anisotropies using similar methodologies. These statistics can be translated into $\Delta\tau$ constraints and compared directly to existing limits on anisotropic optical depth and to $\tau$-reconstruction forecasts. The simulated $C_\ell^{\tau\tau}$, non-Gaussianity, and redshift evolution measured here can also be used to test how well kSZ, 21\,cm, and low-$\ell$ polarization data recover the mean optical depth without assuming an overly simple reionization history. Complementary reionization predictions will naturally incorporate the spatial structure quantified here and help us better understand the high-redshift Universe.

\section*{Acknowledgments}
We thank Cora Dvorkin, Joel Meyers, and Mustapha Ishak for insightful discussions related to this work.
An award of computer time was provided by the INCITE program. This research used resources of the Oak Ridge Leadership Computing Facility at the Oak Ridge National Laboratory, which is supported by the Advanced Scientific Computing Research programs in the Office of Science of the U.S. Department of Energy under Contract No.\ DE-AC05-00OR22725.
The authors acknowledge the MIT Office of Research Computing and Data, FAS Division of Science Research Computing Group at Harvard University, and High Performance Computing at The University of Texas at Dallas (HPC@UTD) for providing resources that have contributed to the research results reported within this paper.
Support for programs JWST-AR-08709 (AS) and JWST-AR-04814 (XS, MV) were provided by NASA through a grant from the Space Telescope Science Institute, which is operated by the Association of Universities for Research in Astronomy, Inc., under NASA contract NAS 5-03127.
Support for OZ was provided by Harvard University through the Institute for Theory and Computation Fellowship.
RK acknowledges support of the Natural Sciences and Engineering Research Council of Canada (NSERC) through a Discovery Grant and a Discovery Launch Supplement (funding reference numbers RGPIN-2024-06222 and DGECR-2024-00144) and York University's Global Research Excellence Initiative.
MV acknowledges support through NASA ATP Grant 23-ATP23-149 and NSF AAG Grant AST-2307699.
VS and LH acknowledge support from the Simons Foundation through the ``Learning the Universe'' initiative.

\bibliographystyle{mnras}
\bibliography{main}

\begin{appendix}

\section{Light-cone Retiling Scale}
\label{ap:retiling}

\begin{figure}
    \centering
    \safeincludegraphics[width=\columnwidth]{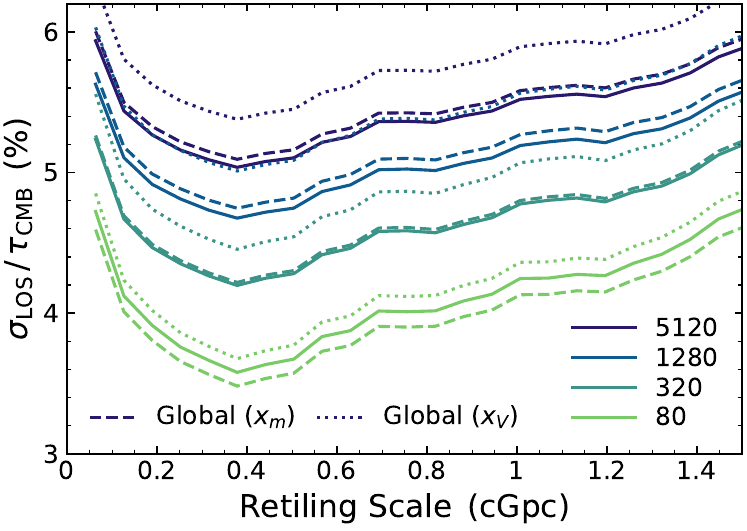}
    \caption{Dependence of the coefficient of variation $\sigma_\text{LOS} / \langle \tau_\text{LOS} \rangle$ on the retiling scale for the light-cone construction, i.e. the fraction of the simulation box length used in each tile before applying rotation and transpose symmetry transformations. While the magnitude of $\sigma/\tau$ is sensitive to the light-cone resolution (the different colors from dark purple to light green correspond to integrating through coarsened representations), the minimum variance tiling of $377$\,cMpc or three-quarters of the simulation box is stable across all resolution levels, and is therefore selected as the fiducial tiling strategy. For completeness, we also show normalizations with respect to $\tau_{\text{CMB},m}$ (dashed) and $\tau_{\text{CMB},V}$ (dotted), which are more sensitive to resolution effects.}
    \label{fig:cadence}
\end{figure}

Given the near alignment of the light cone with one of the simulation axes, visual inspection reveals periodic artifacts emerge from the raw image data. We emphasize that the LOS mean $\langle \tau_\text{LOS} \rangle$ is insensitive to any reordering of pixels in the light cone; however, the variance across sightlines and some other higher-order statistical analyses are affected. A simple but effective choice to mitigate this is to apply the rotation and transpose symmetry transformations to the entire image in an optimal sequence. Specifically, we find cycling through the $D_4$ symmetries of the square to be visually and statistically effective: original orientation, rotated $90^\circ$, rotated $180^\circ$, rotated $270^\circ$, flipped horizontally, flipped vertically, standard transpose, anti-diagonal transpose. We then choose the retiling scale by minimizing the variance across sightlines. In Fig.~\ref{fig:cadence}, we examine how the frequency of retiling the light cone affects the optical-depth statistics. Small retiling scales ($\lesssim 25\%$ of the box) lead to large variance because the same structures are repeated too often, an aliasing effect compounded by the limited set of eight rotations and flips used. Conversely, very large retiling scales ($\gtrsim 100\%$ of the box) also increase variance because the number of independent tiles decreases and extreme nearly-repeated structures can dominate. The variance is minimized around a retiling scale of $377$\,cMpc or $\approx 75\%$ of the box length, which is adopted as the fiducial value in the main analysis as a conservative representation of patchy reionization on the light cone.

\section{Light-cone Resolution}
\label{ap:resolution}

\begin{figure}
    \centering
    \safeincludegraphics[width=\columnwidth]{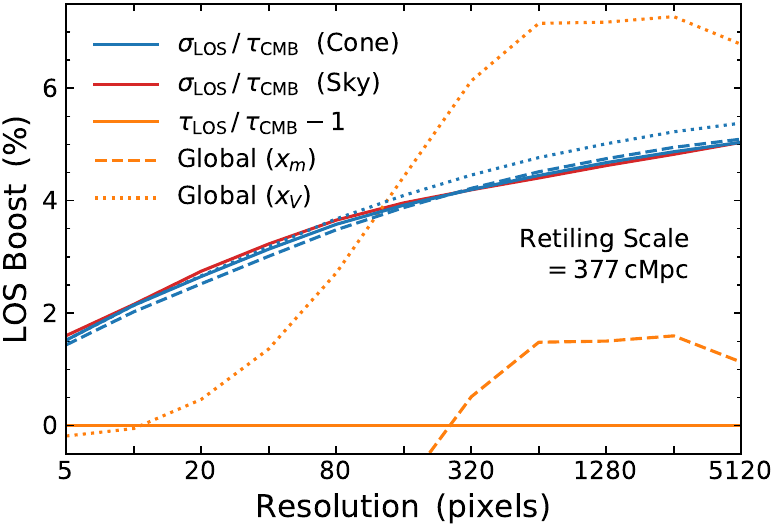}
    \caption{Sensitivity of statistical quantities for $\langle \tau_\text{LOS} \rangle$ maps as a function of coarsening the image resolution before performing sightline integrations, ranging from the fiducial $5120^2$ pixel grid in factors of two down to $5^2$. The orange curves show the relative difference of the LOS and various global averages for each resolution, $\langle \tau_\text{LOS} \rangle / \tau_\text{CMB} - 1$. The solid curve is the self-reference line, the dashed curve adopts a mass-weighted reference $\tau_{\text{CMB},m}$, and the dotted curve uses a volume-weighted reference $\tau_{\text{CMB},V}$, showing the global bias is removed once reaching a resolution of $\sim 500$\,pixels, corresponding to $25''$ or $\sim 1$\,cMpc at $z = 5$. The coefficient of variation $\sigma_\text{LOS} / \langle \tau_\text{LOS} \rangle$ is shown with similarly styled blue curves for spatially-degraded light cones, showing a gradual suppression of variance with coarsening. Finally, the red curve shows $\sigma_\text{LOS} / \langle \tau_\text{LOS} \rangle$ from pure sky degradation of the full resolution image, confirming that there is no systematic bias from the coarsened light-cone integrations.}
    \label{fig:resolution}
\end{figure}

While nearly all of the analysis in this study employed the highest fiducial resolution of $5120^2$ pixels, we briefly assess the impact of utilizing coarsened grids for future analyses, e.g.\ in cases where the large data volume or other computational expenses may not be justified. We degrade the resolution in both spatial and redshift grid dimensions in factors of two down to $5^2$ prior to performing sightline integrations. All LOS analysis is performed with the same retiling scale of $377$\,cMpc. This approach reveals how much small-scale structure contributes to the optical depth. In Fig.~\ref{fig:resolution}, we observe that as the resolution worsens, the coefficient of variation $\sigma_\text{LOS} / \langle \tau_\text{LOS} \rangle$ (blue) decreases as the high-$\tau$ tail is artificially truncated and variance is suppressed. The standard deviation is already a conservative lower limit due to the retiling strategy, but may continue to increase slowly at still higher resolutions. The red curve confirms that there is no systematic bias from the coarsened light-cone integrations compared to pure sky degradation of the full resolution image. However, the LOS mean converges quickly, as shown by the relative difference between the LOS and global averages for each resolution, $\langle \tau_\text{LOS} \rangle / \tau_{\text{CMB},V} - 1$ (orange). Specifically, the LOS boost (relative to volume-weighting) levels off at $\approx 7\%$ once the light-cone maps reach a resolution of $\sim 500$\,pixels, corresponding to $25''$ or $\sim 1$\,cMpc at $z = 5$. Overall, this suggests that the bias introduced by patchiness is not strongly sensitive to the precise resolution of the light cone once bubble-scale structure is resolved. Finally, we add dashed curves to show the impact of replacing the volume-weighted references $\tau_{\text{CMB},V}$ with mass-weighted ones $\tau_{\text{CMB},m}$, confirming that the bias from using the volume-weighted ionization history compared to light-cone sightlines can largely be accounted for by switching to the mass-weighted history.

\section{Diverging Mass and Volume Fractions}
\label{ap:divergence}

\begin{figure}
    \centering
    \safeincludegraphics[width=\columnwidth]{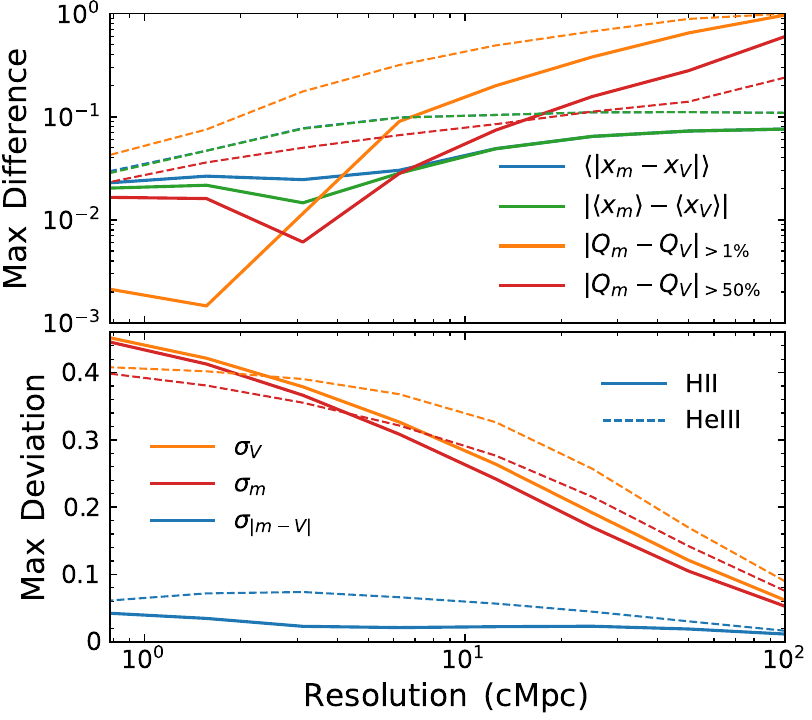}
    \caption{\textit{Top panel:} Resolution dependence of the discrepancy between mass-weighted and volume-weighted ionization fraction fields. For each spatial resolution, we compute four diagnostics plotting the maxima over all redshifts: the mean absolute difference $\langle |x_m - x_V| \rangle$ (blue), the absolute difference of the means $|\langle x_m \rangle - \langle x_V \rangle|$ (green), and the absolute difference in volume-filling fractions $|Q_m - Q_V|$ when the ionization threshold is 1\% (orange) and 50\% (red). \HeIII is shown with thin dashed curves. \textit{Bottom panel:} Standard deviation of the volume-weighted fraction ($\sigma_V$; orange), the mass-weighted fraction ($\sigma_m$; red), and the absolute difference $|x_m - x_V|$ (blue) as a function of resolution. Finer resolutions increase the variance as additional features are better resolved.  Morphological and global statistics are converging on scales $\lesssim 3$\,cMpc, even though small-scale differences remain.}
    \label{fig:Q_max_diff}
\end{figure}

We now examine how well mass-weighted ($x_m$) and volume-weighted ($x_V$) ionization histories agree when the simulation is represented at coarse spatial resolutions. To do so, we utilize the Cartesian data at each coarsened resolution and compute several metrics, summarized by taking the maxima across all redshifts. In the top panel of Fig.~\ref{fig:Q_max_diff}, the blue curve shows the mean absolute difference, $\langle |x_m - x_V| \rangle$, which gradually declines as the resolution improves but remains nonzero even at smaller cell sizes, indicating that unresolved multiphase structure continues to produce some mismatch. The green curve gives the absolute difference of the means, $|\langle x_m \rangle - \langle x_V \rangle|$, which is slightly smaller as global averages converge more quickly than local ones. The orange and red curves provide a more direct morphology measure, as they compare the volume-filling factors of the ionized regions, $|Q_m - Q_V|$, after thresholding at $x>1\%$ and $x>50\%$ ionization levels, respectively. At finer resolutions (cell sizes of $\lesssim 3$\,cMpc), both thresholds show highly suppressed differences, implying that the overall structure of the ionization bubbles is sufficiently well represented. However, when the resolution becomes too coarse ($\gtrsim 10$\,cMpc) the multi-phase averaging leads to diverging mass- and volume-weighted fraction statistics, as expected.

The bottom panel shows the standard deviation of each field as a function of resolution. The orange and red curves track the standard deviations of the volume- and mass-weighted fractions ($\sigma_V$ and $\sigma_m$), which increase at finer resolution because more small-scale substructure becomes resolved. The blue curve gives the standard deviation of $|x_m - x_V|$, which remains small at all resolutions $\lesssim 0.05$ but given the complexity of the bubble size distribution we interpret the local residuals between the two fields as being driven by small, dense features rather than by large-scale trends. \HeIII (thin dashed curves) shows qualitatively similar trends as \HII.

\begin{figure}
    \centering
    \safeincludegraphics[width=\columnwidth]{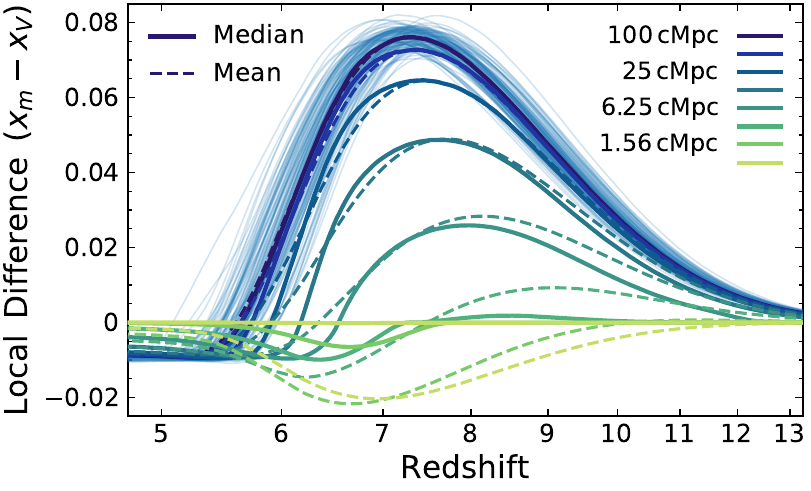}
    \caption{Redshift evolution of the local difference between mass- and volume-weighted ionization fractions ($x_m - x_V$). Thin blue curves represent 125 independent 100\,cMpc subvolumes, tracing the variation in reionization timing across the simulation. For each coarsening scale the thick solid (dashed) curves indicate the median (mean) across all regions. As the cell size decreases from 100\,cMpc (dark purple) to 0.78\,cMpc (light green), skipping every other label for compactness, the peak positive difference at $z \approx 7$ is progressively suppressed and shifts slightly, while the negative dip near $z \approx 5.5$ becomes shallower. The median difference becomes essentially zero at the highest resolutions, indicating that the morphology of ionized regions is converging at $\lesssim 3$\,cMpc. The mean difference remains mildly negative when self-shielded clouds reside in volume-ionized regions.}
    \label{fig:x_diff_5}
\end{figure}

Using the maximum as the summary statistic is a conservative choice and other metrics such as mean and median lead to qualitatively similar behavior but with smaller values, reflecting other aspects of the time evolution. To illustrate this point, in Fig.~\ref{fig:x_diff_5}, we show the local difference ($x_m - x_V$) as a function of redshift, and how it depends on the coarsening scale. The thin blue curves are 125 independent 100\,cMpc subvolumes, highlighting that the individual differences are driven by the variation in the timing of reionization across the simulation, mirroring the similar evolution of the global difference shown in Fig.~\ref{fig:x_boost}. The thick curves collapse the detailed variation into the median (solid) and mean (dashed) statistics, showing that these continue to decrease as the resolution increases from 100\,cMpc to 0.78\,cMpc. As before, the median is significantly suppressed below $\lesssim 3$\,cMpc and is indistinguishable from the zero line at 0.78\,cMpc. The mean remains negative due to unresolved self-shielded substructure that will likely persist to some degree in the differences down to the resolution limit of the simulation, and in reality below that. Overall, while local differences persist, the mean ionization history and the morphology of ionized regions are converging once the simulation resolves scales of order a few cMpc, lending confidence to the analysis of $\tau_\text{CMB}$ and related statistics. These conclusions apply to post-processing IGM analyses and do not imply that RHD simulations would be accurate at such coarse resolutions.

\end{appendix}

\end{document}